\documentclass[10pt, sigconf]{acmart}
\usepackage{hyperref}
\usepackage{enumitem}
\usepackage{listings}
\usepackage{cleveref}
\usepackage{booktabs} 
\usepackage{xspace}
\usepackage{xcolor}
\usepackage{microtype}
\usepackage{subcaption}
\usepackage{balance}

\definecolor{darkgreen}{rgb}{0.15,0.55,0.15}
\definecolor{darkblue}{rgb}{0.1,0.1,0.5}
\definecolor{blue}{rgb}{0.19,0.58,1}
\definecolor{darkgreen}{rgb}{0.15,0.55,0.15}
\definecolor{mred}{rgb}{.80,.12,.30}
\definecolor{grey}{rgb}{0.5,0.5,0.5}
\definecolor{Purple}{rgb}{.75,0,.85}
\definecolor{light-gray}{gray}{0.95}
\definecolor{mid-gray}{gray}{0.85}
\definecolor{darkred}{rgb}{0.7,0.25,0.25}
\newcommand{\red}[1]{\textcolor{red}{#1}}

\newcommand{\blue}[1]{\textcolor{blue}{#1}}

\newcommand{\eat}[1]{}

\newcommand{\stitle}[1]{\vspace{2pt}\noindent\textbf{#1}}

\newlength{\listingindent}                
\usepackage[LGRgreek]{mathastext}

\newtheorem{defi}{Definition}

\newcommand{\sys}[0]{DeepBase\xspace}

\newcommand{\madlib}[0]{\texttt{MADLib}\xspace}
\newcommand{\pybase}[0]{\texttt{PyBase}\xspace}
\newcommand{\naive}[0]{\texttt{+MM}\xspace}
\newcommand{\cpu}[0]{\texttt{+MM (CPU)}\xspace}
\newcommand{\gpu}[0]{\texttt{+MM (GPU)}\xspace}

\newcommand{\appr}[0]{\texttt{+MM+ES}\xspace}

\renewcommand\footnotetextcopyrightpermission[1]{} 
\begin{document}

\title{\sys: Deep Inspection of Neural Networks}
\subtitle{Technical Report}

\settopmatter{authorsperrow=4}

\author{Thibault Sellam}
\affiliation{\institution{Columbia University}}
\email{tsellam@cs.columbia.edu}

\author{Kevin Lin}
\affiliation{\institution{Columbia University}}
\email{kl2806@columbia.edu}

\author{Ian Huang}
\affiliation{\institution{Columbia University}}
\email{iyh2110@columbia.edu}

\author{Yiru Chen}
\affiliation{\institution{Columbia University}}
\email{yc3515@columbia.edu}

\author{Michelle Yang}
\affiliation{\institution{UC Berkeley}}
\email{michelleyang@berkeley.edu}

\author{Carl Vondrick}
\affiliation{\institution{Columbia University}}
\email{cv2428@columbia.edu}

\author{Eugene Wu}
\affiliation{\institution{Columbia University}}
\email{ewu@cs.columbia.edu}

\renewcommand{\shortauthors}{T. Sellam et al.}
\begingroup
\mathchardef\UrlBreakPenalty=10000

\begin{abstract}
  Although deep learning models perform remarkably well across a range of tasks such as language translation and object recognition, it remains unclear what high-level logic, if any, they follow. Understanding this logic may lead to more transparency, better model design, and faster experimentation. Recent machine learning research has leveraged statistical methods to identify hidden units that behave (e.g., activate) similarly to human understandable logic, but those analyses require considerable manual effort. Our insight is that many of those studies follow a common analysis pattern, which we term Deep Neural Inspection.  There is opportunity to provide a declarative abstraction to easily express, execute, and optimize them.

This paper describes \sys, a system to inspect neural network behaviors through a unified interface. We model logic with user-provided \emph{hypothesis functions} that annotate the data with high-level labels (e.g., part-of-speech tags, image captions). DeepBase lets users quickly identify individual or groups of units that have strong statistical dependencies with desired hypotheses. We discuss how \sys can express existing analyses, propose a set of simple and effective optimizations to speed up a standard Python implementation by up to $72\times$, and reproduce recent studies from the NLP literature.
\end{abstract}

\maketitle

\section{Introduction}
\label{s:intro}

Neural networks (NNs) are revolutionizing a wide range of machine intelligence tasks with impressive performance, such as language understanding~\cite{graves2014towards}, image recognition~\cite{girshick2014rich}, and program synthesis~\cite{devlin2017robustfill}. This progress is partly driven by the proliferation of deep learning libraries and programming frameworks that drastically reduce the effort to construct, experiment with, and deploy new models~\cite{baylor2017tfx,crankshaw2014missing,kumar2017data}.

However, it is still unclear how and why neural networks are so effective~\cite{doshi2017towards,nytcats}.  Does a model learn to decompose its task into understandable sub-tasks? Does it memorize training examples~\cite{zhang2016understanding}, and can it generalize to new situations?
Grasping the internal representation of neural networks is currently a major challenge for the machine learning community. While the field is still in its infancy, many hope that increasing our understanding of trained models will enable more rapid experimentation and development of NN models, help identify harmful biases, and explain predictions~\cite{doshi2017towards}---all critical in real-world deployments.

A prevailing paradigm is to study how individual or groups of hidden units (neurons) behave when the model is evaluated over test data.  One approach is to identify if the behavior of a hidden unit mimics a high level functionality---if a unit only activates for positive product reviews, then it potentially recognizes positive sentiment.  Numerous papers have applied these ideas by manually inspecting visualizations of behaviors~\cite{karpathy2015visualizing,radford2017learning,girshick2014rich} or writing analysis-specific scripts~\cite{shi2016emnlp,bau2017network}, and in domains such as detecting syntax and sentiment in language~\cite{karpathy2015visualizing,radford2017learning}, parts and whole objects from images~\cite{girshick2014rich,le2013building}, image textures~\cite{bau2017network}, and chimes and tunes in audio~\cite{aytar2016soundnet}.

This class of analysis is ubiquitous in the deep learning literature~\cite{noroozi2016unsupervised,alain2016understanding,belinkov2017neural,kim2017tcav,kadar2017representation,bau2017network,morcos2018importance}, it is particularly well represented in neural net interpretability workshops~\cite{blackbox}, yet each analysis is implemented in an ad-hoc, one-off basis.  In contrast, we find that they belong in a common class of analysis that we term \textbf{\textit{Deep Neural Inspection (DNI)}}.  Given user-provided hypothesis logic (e.g., ``detects nouns'', ``detects keywords''), DNI seeks to quantify the extent that the behavior of hidden units (e.g., the magnitude or the derivative of their output) is similar to the hypothesis logic when running the model over a test set. These DNI analyses share a common set of operations, yet each analysis currently requires considerable engineering (hundreds or thousands of lines of code) to implement, and often runs inefficiently.  \emph{We believe there is tremendous opportunity to provide a declarative abstraction to easily express, execute, and optimize DNI analysis.}

Our main insight is that DNI analyses primarily use statistical measures to quantify the affinity between hidden unit behaviors and hypotheses, and simply differ in the specific NN models, hypotheses, or types of hidden unit behaviors that are studied.  \Cref{ss:motivation} illustrates how existing DNI analyses fit this pattern~\cite{alain2016understanding,karpathy2015visualizing,zhou2017interpreting}.  To this end, we designed \sys, a system to perform large-scale Deep Neural Inspection through a declarative interface. Given groups of hidden units, hypotheses, and statistical affinity measures, \sys quickly computes the affinity score between each {\it(hidden units group, hypothesis)} pair. The aim is for \sys to accelerate the development and usage of this class of neural network analysis in the ML community.

Designing a fast DNI system is challenging because the cost is cubic with respect to the size of the dataset, the number of hidden units, and the number of hypotheses.  Even trivial examples are computationally expensive.  Consider analyzing a character-level recurrent neural net (RNNs) with 128 hidden units over a corpus of 6.2M characters~\cite{karpathy2015visualizing}.  Assuming each activation is stored as a 4-byte float, each RNN model requires 3.1GB to store its activations.   While this fits in memory, the process of extracting these activations, storing them, and matching them with hundreds of hypotheses can be incredibly slow.

To address this challenge, \sys uses pragmatic optimizations. First, users provide hypothesis logic as functions evaluated over input data, which may be computationally expensive.  \sys can cache the output of hypothesis functions and reuse their results when re-running the same DNI analysis on new models.
Second, users can specify convergence thresholds so that \sys can terminate quickly while returning accurate but approximate scores. \sys natively provides popular measures such as correlation and linear prediction models.
Third, \sys reads the dataset, extracts unit behaviors, and evaluates the user-defined hypothesis logic in an online fashion, and can terminate the moment the affinity scores have converged.
Finally, \sys leverages GPUs---commonplace in deep learning---to offload and parallelize the costs of extracting unit behaviors and computing affinity metrics based on e.g., logistic regression.

\textbf{Our primary contribution is to formalize Deep Neural Inspection and develop a declarative interface to specify DNI analyses}.  We also contribute:
\begin{itemize}[leftmargin=*, topsep=0mm, itemsep=0mm]
    \item The design and implementation of an end-to-end DNI system called \sys, along with simple pragmatic optimizations, including caching, early stopping via convergence criteria, streaming execution, and GPU execution.
    \item A walk-through of how to generate hypothesis functions from existing ML libraries.
    \item Extensive performance experiments based on a SQL auto-completion RNN model.  We show that with all optimizations including caching, \sys outperforms a standard Python baseline by up to 72X, and an in-RDBMS implementation using MADLib~\cite{hellerstein2012madlib} by $100-419\times$, depending on the specific affinity measure.
    \item Experimental results using \sys to analyze a state-of-the-art Neural Machine Translation model architecture~\cite{opennmt} (English to German). We compare \sys to existing scripts~\cite{belinkov2017neural} and validate the results of recent NLP research~\cite{shi2016emnlp, adebayo2018local}.
\end{itemize}

\noindent This paper focuses the discussion and application of \sys on Recurrent Neural Networks (RNNs), a widely used class of NNs used for language modeling, program synthesis, image recognition, and more.  We do this to simplify the exposition while focusing on an important class of NNs, however \sys also Convolutional Neural Networks (CNNs).  See \Cref{a:dni-cnn} for a results comparison with a recent system called NetDissect~\cite{zhou2017interpreting}, and our prior work for more CNN and Reinforcement Learning examples~\cite{dninips18}.
\section{Background and Use Cases}
\label{s:background}

This section introduces current examples of DNI analyses and how they are implemented. These uses cases serve as the motivation for the system described in the rest of the paper.
Appendix~\ref{sec:primer} provides a quick primer on Neural Networks (NNs) and explains the terminology.   The important concept is that a NN is composed of hidden units, and when a NN is evaluated over an input record (e.g., a sequence of characters that form a sentence, or a matrix of pixels that form an image), each hidden unit performs an action that emits a behavior value (e.g., an activation, or the derivative of an activation) for each element of the record (e.g., character or pixel).  We refer the reader to Appendix~\ref{sec:primer} for details.

\subsection{Motivating Example}
\label{ss:motivation}

We use a recurrent model that performs SQL query auto-completion as a motivating example. Given a SQL string, the model can read a window of 100 characters (padded if necessary) and predict the character that follows. Technically, the neural net comprises three layers: one input layer that reads one-hot-encoded characters, one recurrent (LSTM) layer with 500 hidden units, and one fully connected layer for the final output. The analysis focuses on the recurrent layer.

The model achieve 80\% prediction accuracy on a held-out test set of 1,152 queries, as compared with random guess accuracy of $\frac{1}{32}$. The statistics indicate that the model can reliably predict the next character, {\it but what did the model really learn?} One hypothesis is that the model ``memorized'' all possible queries. Another is that it learns an $N$-gram model that uses the previous $N-1$ characters to predict the next. Or the model learned portions of the SQL grammar, e.g., it learned that column references tend to follow the \texttt{SELECT} keyword and that table names usually come after \texttt{FROM} but not before \texttt{LIMIT}. Ideally, we would also like to check these hypotheses across models with different architectures or training parameters, or for a specific set of hidden units.

\subsection{Approaches for Interpretation}
\label{sec:approaches}
The machine learning community has developed a variety of approaches for interpretation, which we discuss below.

\label{sec:example}
\begin{figure}[bt]
    \includegraphics[width=\columnwidth]{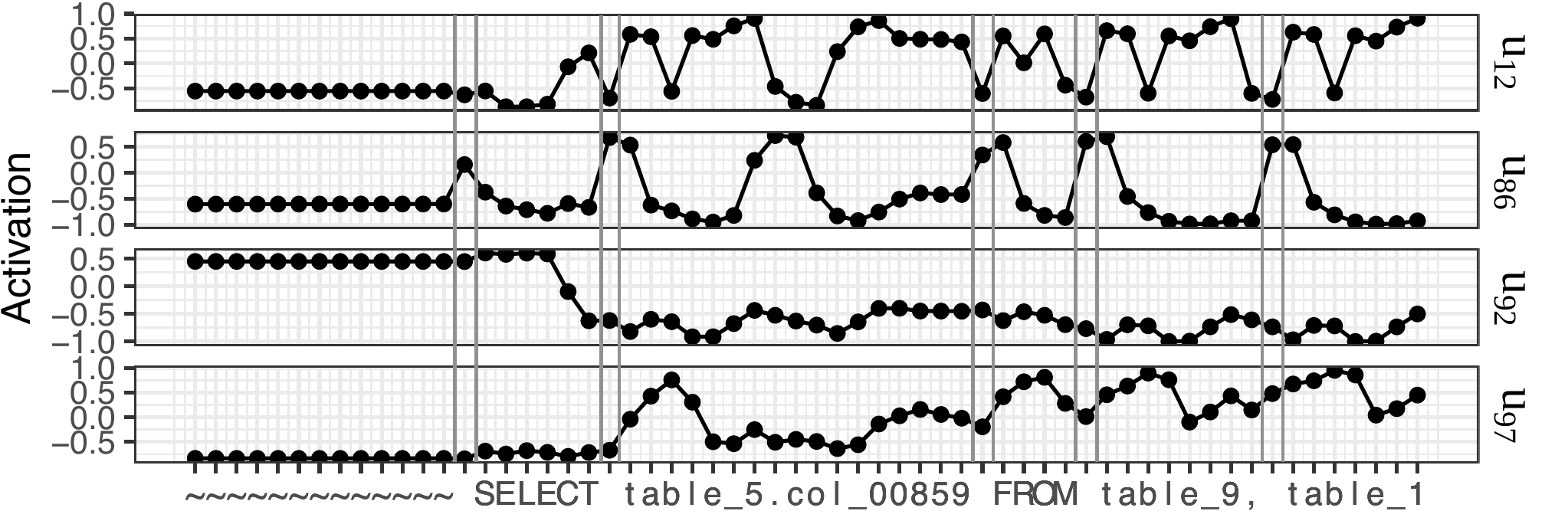}
    \vspace{-.1in}
    \caption{Activations over time for the SQL auto-completion model. What is the model learning?}
    \vspace{-.1in}
    \label{f:activations}
\end{figure}

\stitle{Manual Visual Inspection:} Manual approaches~\cite{strobelt2018lstmvis,karpathy2015visualizing}, such as LSTMVis~\cite{strobelt2018lstmvis}, visualize each unit's activations and let users manually check that the units behave as expected. For instance, if a unit only spikes for table names, it suggests that the model behaves akin to that grammar rule and possibly has ``learned'' it.
Unfortunately, visual inspection can turn out to be challenging, even for simple settings. Figure~\ref{f:activations} plots the activations of 4 units on the prefix of a query.  We easily observe that units are inactive when reading the padding character ``\texttt{\~}''. However, interpreting the fluctuations is difficult. $u_{12}$ appears to spike down on whitespaces (highlighted), mirrored by $u_{86}$. $u_{97}$ tends to activate within the words \texttt{FROM} and \texttt{table}.  But those observations are simply guesses on a small string; scaling this manual analysis to all units and all queries is impractical.  Ideally, we would express these hypotheses and formally test them at scale.

\stitle{Saliency Analysis: } This approach seeks to identify the input symbols that have the largest ``effect'' on a single or group of units.  For instance, an NLP researcher may want to find words that an LSTM's output is sensitive to~\cite{li2015visualizing}, or the image pixels that most activates a unit~\cite{girshick2014rich}.  This analysis may use different behaviors, such as the unit activation or its gradient. Typically, the procedure collects a unit's behaviors, finds the top-k highest value behaviors, and reports the corresponding input symbols. For instance, whitespaces and periods trigger the five highest activations for $u_{86}$ in \Cref{f:activations}.  This DNI approach has been used to analyze image object detection~\cite{simonyan2013deep,selvaraju2016grad,zhou2014object}, in NLP models~\cite{li2015visualizing} and sentiment analysis~\cite{radford2017learning}.

\stitle{Statistical Analysis: } Many datasets are annotated: text documents are annotated with parse trees or linguistic features, while image pixels are annotated with object information. Such annotations can help analyze groups of units.

In our SQL example, we could parse the query and annotate each token with the name of its parent  rules (e.g., \texttt{where\_clause} or \texttt{variable\_name}). If we find a strong correlation between the activations of a hidden unit and the occurrence of a particular rule while running the model (e.g., ``hidden unit 99 has a high value for every token inside \texttt{WHERE} predicates''), then we have some evidence that the hidden unit acts as a detector for this rule~\cite{karpathy2015visualizing}. We could take the analysis further and test \emph{groups} of hidden units: if we build a classifier on top of their activations and find that it can predict the occurrence of grammar rules with high accuracy, then we have evidence that those neurons behave \emph{collectively} as a detector~\cite{alain2016understanding,belinkov2017neural}.

Statistical analysis of hidden unit activations is a widespread practice in the machine learning literature. For instance, Kim et. al~\cite{kim2017tcav} use logistic regression to predict annotations of high-level concepts from unit activations.  NetDissect~\cite{bau2017network} finds the image pixels that cause a unit to highly activate (similar to saliency analysis), and computes the Jaccard distance between those pixels and annotated pixels of e.g., a dog.  In general, these techniques compute a statistical measure between unit behaviors and annotations of the input data, and have been used to e.g., find semantic neurons~\cite{morcos2018importance}, compare models~\cite{raghu2017svcca} or more generally evaluate to what extent neural nets learn high-level concepts such as textures or part-of-speech tags~\cite{noroozi2016unsupervised,alain2016understanding,belinkov2017neural, bau2017network}.

\stitle{Inspection in Practice:} Although the model interpretation literature is extremely active,
the software ecosystem of tools to support Deep Neural Inspection is very limited. Authors have focused on reproducibility in the narrow sense, rather than usability, and it is likely that a ML engineer will have to implement her own version of a given approach\footnote{These remarks don't apply to the NN visualization community, which publishes and maintains several important software packages~\cite{strobelt2018lstmvis, olah2018the}.}.

We searched online for software used in publications that perform deep neural inspection~\cite{belinkov2017neural,kadar2017representation,radford2017learning,shi2016emnlp,bau2017network, kim2017tcav,zhou2014object,noroozi2016unsupervised,alain2016understanding,morcos2018importance,zeiler2014visualizing}.  Of these, six papers provided code repositories and four of them target computer vision models. In all cases, the scripts (in various languages) are tailored to only reproduce the experiment described the corresponding papers---that is, they are custom implemented for one type of model, one type of analysis, and for one type of dataset. All scripts have different APIs, and several rely on outdated/unsupported versions of deep learning frameworks (LuaTorch, PyTorch, Caffe, or Tensorflow). Popular approaches such as~\cite{zeiler2014visualizing} also have ``unofficial'' implementations that exhibit similar issues.

\begin{figure}
    \vspace{-.1in}
    \includegraphics[width=.7\columnwidth]{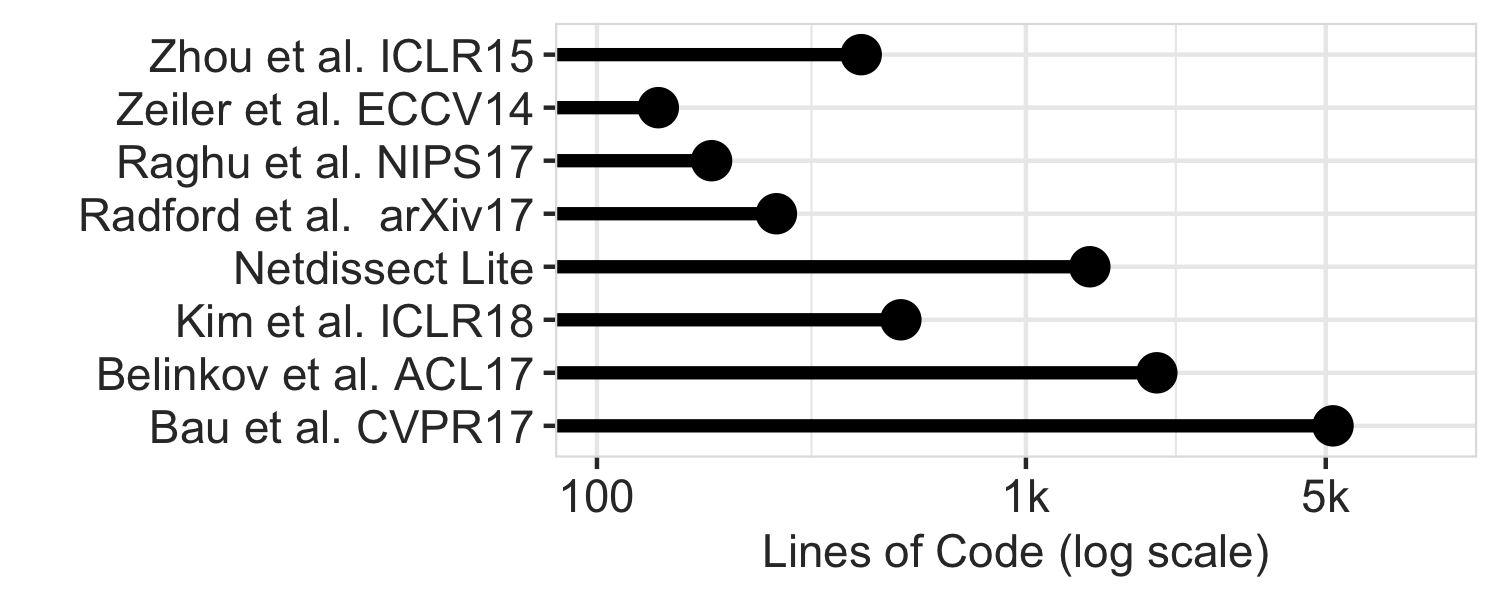}
    \caption{Lines of code (approx) from available code for papers that perform DNI.}
    \vspace{-.1in}
    \label{f:stats_dnipapers}
\end{figure}

Figure~\ref{f:stats_dnipapers} summarizes the lines of code in each repository after manually removing non-essential code (e.g., non-analysis visualization or imported libraries). Every analysis is at least several hundred lines of code, and in some cases thousands of lines. Although this is an imperfect measure, it provides a sense of the complexity of current DNI methods.

\subsection{Desiderata of a DNI System }

DNI analysis using the existing approaches is powerful and spans many domains of application. Unfortunately, each analysis currently requires custom, ad-hoc implementations despite following a common analysis goal: \textbf{the user wants to measure the extent that groups of hidden units in one or more trained models behave in a manner that is similar to, or indicative of, a human-understandable function, when evaluated over the same test dataset}.    For instance, we may want to measure to what extent the activations of each hidden unit in our SQL auto-completion model correlates with the output of a function that detects the presence SQL keywords by emitting $1$ for keyword characters and $0$ otherwise.

\sys is a system that provides a declarative abstraction to efficiently express and execute these analyses. \sys takes as input a test set, a trained model, a set of Python functions that encode hypotheses of what the model may be learning (we call them \emph{hypothesis functions}), and a scoring function, e.g., a measure of statistical dependency. From those inputs, \sys{} produces a set of scores that quantify the affinity between the hypotheses and the model's hidden units. Such a system should support:

\stitle{Arbitrary Hypothesis Logic:} Different applications and domains care about different hypotheses.  In auto-completion, does the model learn to count characters?  In machine translation, do units learn sentiment or language nuances such as relational dependencies? In visual object recognition, pixels correspond to different types of objects---do units detect pixels containing dogs or cats? A system should be flexible about the types of logic that can be used as queries.

\stitle{Many Models and Units:} Modern neural network models can contain tens of thousands of hidden units, and researchers may want to compare across different model architectures, training epochs, parameters, datasets, or groups of units.  A DNI system should allow users to easily specify which combination of hidden units and models to inspect.

\stitle{Different Affinity Measures:} Different use cases and users may define affinity differently.  They may correlate activations of individual units~\cite{karpathy2015visualizing}, compute mutual information between a unit's behavior and annotations~\cite{morcos2018importance}, use a linear model to predict high-level logic from unit activations~\cite{alain2016understanding,kim2017tcav}, or use another measure.  A DNI system should be fast for common measures, and support user-defined measures.

\stitle{Mix and Match: }  Users should be able to easily specify the combination of hypothesis functions, models, hidden units, and datasets that they want to inspect.

\stitle{Analyze Quickly: } Developers use inspection functionality to interactively debug and understand the characteristics of their models.  Thus, any system should both scale to a large number of models, test data, and queries, while maintaining acceptable query performance.

\section{Problem Definition}\label{s:probdef}

We now define the deep neural inspection problem, using the SQL auto-completion model in \Cref{ss:motivation} as the example.

\stitle{Problem Setup: }  Let a dataset $D$ be a $n_d\times n_s$ matrix of symbols where $d_i$ is the $i^{th}$ row (or record) of size $1\times n_s$ (Table~\ref{tbl:notation} presents our notations).  In our SQL auto-completion example, each record is a $100$-symbol vector where each symbol is a one-hot encoded character\footnote{Each character is represented by a sparse binary vector, where a 1 at position $\emph{i}$ indicates that the character is set to the $\emph{i}$th value from the alphabet.}.  For other data types, a symbol may be an image pixel, a word, or a vector depending on the model.  Records are null-padded to ensure that all records are the same size.

A model $M$ is a vector of $n_M$ hidden units, where $u^M_i$ is the $i^{th}$ unit.  Logically, $M(d)$ is evaluated on a record $d$ by reading each input symbol one at a time; each symbol $s_i$ triggers a single behavior $b_i\in \mathbb{R}$ from each hidden unit $u$\footnote{In the context of windowing over streaming data,  the RNN model internally encodes a dynamically size sliding window over the symbols seen so far.}.  Thus the {\it Unit Behavior} $u(d)\in\mathbb{R}^{n_s}$ is the vector of behaviors for unit $u$ when the models evaluated over all symbols in $d$. Let $U(d)\in\mathbb{R}^{|U|\times n_s}$ be the {\it Group Behavior} for a subset of units $U$ in a model.  An example of $U$ may be the units in the first layer, or simply all units in a model.

Each line graph in \Cref{f:activations} plots behavior as a unit's activation when reading each character in the input query.  This paper reports results based on unit activations, however \sys is agnostic to the specific definition of behavior extracted from the model.  This flexibility is important because some papers use the gradient of the activations instead of their magnitude~\cite{zhou2014object}.

We model high level logic in the form of a {\it Hypothesis Function}, $h(d)\in \mathbb{R}^{n_s}$, that outputs a {\it Hypothesis Behavior} when evaluated over $d$. In practice, those functions are either written by the user or provided in a library.  There is no restriction on the complexity of a hypothesis function and \Cref{ss:hypotheses} describes example functions; the only constraint is that the hypothesis behavior is size $n_s$ so that it matches the size of a unit behavior.

To illustrate, a hypothesis that the model has learned to detect the keyword ``\texttt{SELECT}'' could be a Python function that emits $1$ for those characters and $0$ otherwise.  Thus for the query \texttt{SELECT 1 FROM a}, the hypothesis would be \texttt{111111000000000}.  The hypothesis behavior need not be binary, and can encode integers or floating point values as well.  For instance, a hypothesis that the model counts the number of characters in an input string may return a number between 0 an 100.  Further examples are given in \Cref{ss:hypotheses}.
\begin{table}[t!]
\centering
{\small
\begin{tabular}{rl}
    \textbf{Notation} & \textbf{Description}\\
    \midrule
    $D$   & A $n_d\times n_s$ matrix of symbols. \\
    $d_i$  & The $i^{th}$ row in D.  Called a record. \\
    $M$    & A model $M$ is a vector of hidden units. \\
    $u^M_i$    & $i^{th}$ hidden unit in $M$. \\
    $U$ & A group of units in a model. \\
    $u(d)\in\mathbb{R}^{n_s}$  & Unit $u$ returns vector of behaviors.\\
    $h(d)\in\mathbb{R}^{n_s}$  & Hypothesis function $h$ returns vector of behaviors.\\
    $n_d, n_s, n_M$ & Number of records, symbols/record, units in $M$\\
\end{tabular}
}
\caption{Summary of notations used in the paper. }
\label{tbl:notation}
\end{table}

We quantify the affinity between  a group of units $U$ and a hypothesis $h$ with a user-defined statistical affinity measure $l(U, h, D) = (\mathbb{R}^{|U|}, \mathbb{R})$.  The first element contains a scalar affinity score for each unit in the group, and the second element is a score for the group as a whole.  Either element may be empty. For instance, we may replicate~\cite{karpathy2015visualizing} by computing the correlation each unit's activation and a grammar rule such as ``\texttt{SELECT}'' keyword detection.  Alternatively, we may follow~\cite{belinkov2017neural} and use a linear classifier to predict the occurrence of the keyword from the behavior of all units in the first layer; the model's F1 score is the group affinity, and each unit's score is its model coefficient.  Although $l(U, h, D)$ is user-defined, \sys provide 8 common measures (see \ref{sec:measures}) and leverage their approximation properties to optimize the analysis runtime (Section~\ref{s:opts}).

\stitle{Basic Problem Definition: } Given the above definitions, we are ready to define the basic version of DNI:
\begin{defi}[DNI-Basic]\label{p:basic}
  Given dataset $D$, a subset of units $U\subseteq M$ of an RNN model $M$, hypothesis $h$, statistical measure $l$, return the set of tuples $(u, s_u, s_U)$ where the score $s_u$ is defined as $l(U, h, D) = ([s_u | u \in U], s_U)$.
\end{defi}
Note that we specify as input a set of units $U$ rather than the full model $M$.  This is because the statistical measure $l()$ may assign different affinity scores depending on the group units that it analyzes.  For instance, if the user inspects units in a single layer using logistic regression, then only the behaviors of those units will be used to fit the linear model and their coefficients will be different than when inspecting all in the model.   This highlights the value of embedding \sys within a SQL-like language.

\stitle{General Problem Definition: } Although the above definition is sufficient to express the existing approaches in \Cref{ss:motivation}, it is inefficient. In practice, developers often train and compare many groups of units, e.g., to understand what hypotheses the model learns across training epochs.  We present a more general definition that is amenable to optimizations across models, hypotheses, and measures.

Let $\mathbb{U}$ be a set of unit groups defined by the user. The user may also provide a large corpus of hypotheses $H$, to understand which hypotheses are learned by the model. The user may also want to evaluate multiple statistical measures $L$ to have different perspectives. With those notations, we define our problem as follows:

\begin{defi}[DNI-General]\label{p:general}
  Given dataset $D$, set of unit groups $\mathbb{U}$, hypotheses $H$,
  and measures $L$,
  return the set of tuples $(u, h, l, s_{u,h,l}, s_{U,h,l})$ where
  \begin{itemize}[topsep=0mm, itemsep=0mm]
    \item $l(U, h, D) = ([s_{u,h,l} | u \in U ], s_{U,h,l})$
    \item $l \in L$, $U \in \mathbb{U}$, $h \in H$
  \end{itemize}
\end{defi}

\section{\sys API and Overview}
\label{sec:overview}

This section describes our Python API, how to create hypothesis functions, \sys's native affinity measures, and a verification procedure to assess the quality of highly scored units.  The next section describes the system design and optimization.

\subsection{Python API}
\label{sec:api}

\sys is implemented in Python and exposes a Python API. We will use the API to perform two analyses using  the SQL-autocompletion example: 1) compute the correlation between every unit's activations and binary hypotheses that indicate the occurrence of grammar rules (as described in Section~\ref{sec:approaches}), and 2) report the F1 accuracy of a logistic regression classifier that predicts the binary hypothesis behaviors from all hidden unit activations~\cite{alain2016understanding, belinkov2017neural}:

{\small\begin{verbatim}
import deepbase
model   = load_model('sql_char_model.h5')
dataset = load_data('sql_queries.tok')
scores  = [CorrelationScore('pearson'),
           LogRegressionScore(regul='L1',score='F1')]
hypotheses = gram_hyp_functions('sql_query.grammar')
deepbase.inspect([model], dataset, scores, hypotheses)
\end{verbatim}}

\smallskip
This code loads the \texttt{deepbase} module, NN model, and test dataset.  It specifies that we wish to compute per-unit correlation scores as well as logistic regression F1 accuracy with L1 regularization.  \texttt{hypotheses} is a list of binary hypothesis functions that each returns the presence of a specific grammar rule.  Finally, we call \texttt{deepbase.inspect()}, which returns a Pandas data frame (i.e., table) that contains an affinity value for each model, score, hypothesis, and hidden unit:
{\small\begin{verbatim}
    model_id, score_id, hyp_id, h_unit_id, val
\end{verbatim}}
\noindent The variable \texttt{scores} points to a list of \texttt{DBScores} objects. Currently, \sys's standard library includes 8 scores (see \ref{sec:measures}) and 2 naive baselines (random class, majority class).  The list \texttt{hypotheses} contains arbitrary Python functions, which output formats are checked during execution (we defined the specifications in \ref{s:probdef}).

In practice, the users will often post-process the table returned by \texttt{inspect}. For instance,
they may wish to return only the top scores (e.g., to find the ``sentiment neuron'' in \cite{radford2017learning}), combine the results with other statistics (e.g., to reproduce Figure 2 in \cite{belinkov2017neural}), or group the scores by layer and count the number of hidden units with a high score (Figure 5 in \cite{bau2017network}). This observation, combined with the fact that the intermediate and final outputs can be very large (multiple GBs for even simple cases) calls for tight integration with a DBMS.  A full treatment is outside the scope of this paper, however Appendix~\ref{ss:inspect} describes how SQL can be extended to support DNI using a new \texttt{INSPECT} clause.  In addition, Section~\ref{sec:naive} describes a baseline built upon a database engine rather than the Python and Tensorflow scripts used in existing papers.

\subsection{Hypotheses}\label{ss:hypotheses}

Hypotheses are the cornerstone of DNI analyses, as they encode the logic that we search for.
Although numerous language-based models, grammars, parsers, annotations, and other information already exist, many do not fit the hypothesis function abstraction.  For example, parse trees (Figure~\ref{f:parsing}) are a common representation of an input sequence that characterizes the roles of different subsequences of the input.  What is an appropriate way to transform them into hypothesis functions?

This section provides examples for generating hypothesis functions from common machine learning libraries that we used in our experiments.
We note that the purpose of \sys is to simplify the use and inspection of hypothesis functions---developing appropriate hypothesis functions to answer NN analysis questions continues to be an open area of research.

\begin{figure}
  \centering
  \vspace{-.1in}
  \includegraphics[width=0.7\columnwidth]{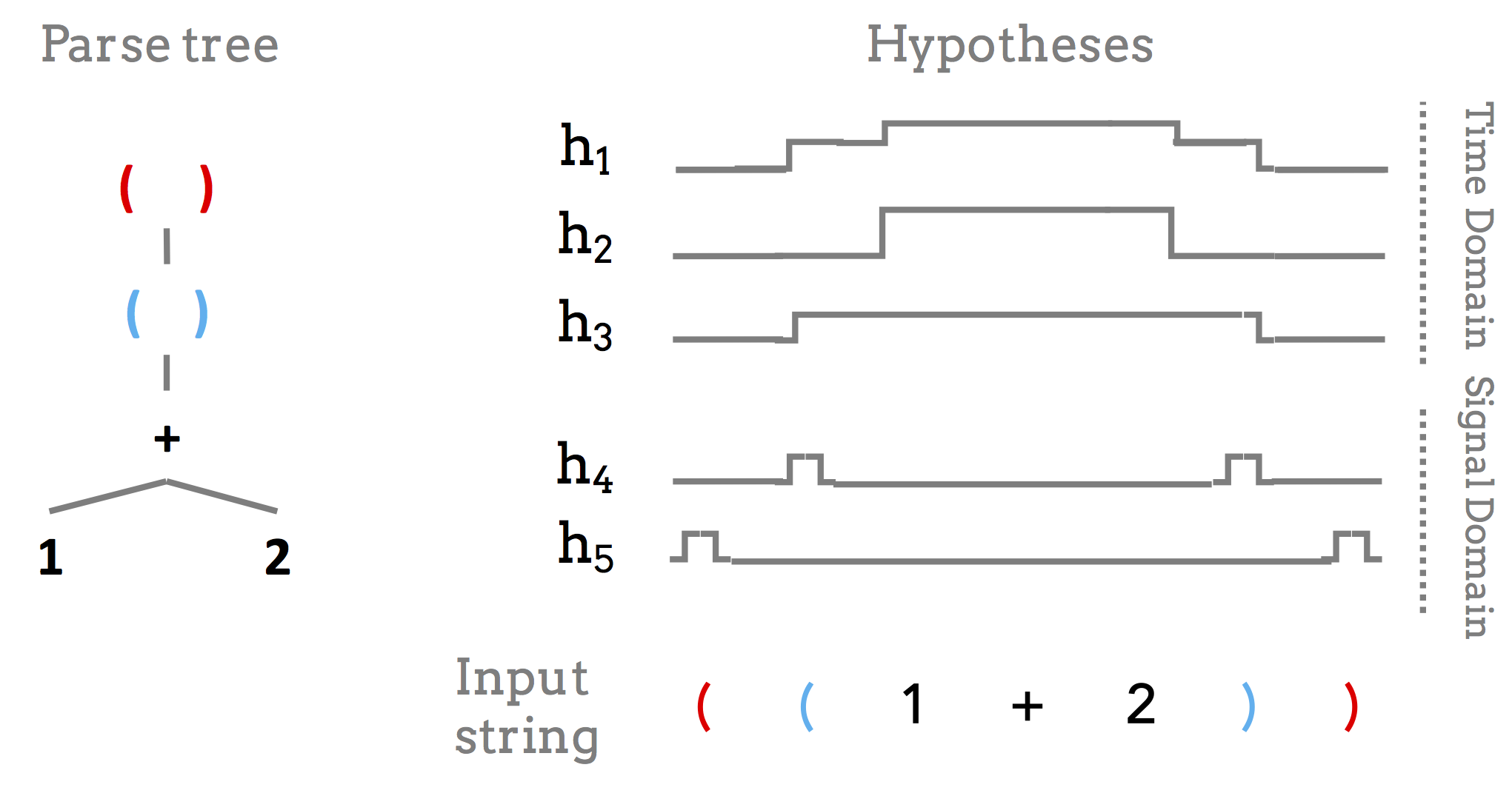}
  \vspace{-.1in}
  \caption{Example parse tree (left) and hypothesis functions.  Each $h_i$ is the behavior over each input character symbol. }
  \label{f:parsing}
\end{figure}

\stitle{Parse Trees: }
A common use of RNNs is language analysis and modeling. For these applications, there are decades of research on language parsing, ranging from context free grammars for programming languages to dependency and constituency parsers for natural language.  Figure~\ref{f:parsing} illustrates an example parse tree for a simple algebraic expression within nested parentheses \texttt{((1+2))}.   The corresponding parse tree contains leaf nodes that represent characters matching terminals, and intermediate nodes that represent non-terminals.

Given a parse tree, we map each node and node type to a hypothesis function.  To illustrate, the \red{red} root in Figure~\ref{f:parsing}'s parse tree corresponds to the outer \red{\texttt{()}} characters.  It can be encoded as a time-domain representation that activates throughout the characters within the parentheses ($h_3$), or a signal representation  that activates at the beginning and end of the parentheses ($h_5$).  Similarly, $h_2$ and $h_4$ are time and signal representations generated by hypothesis functions for the inner \blue{blue} parentheses. Finally, $h_1$ is a composite of $h_2$ and $h_3$ that accounts of the nesting depth for the parentheses rule. Note that a given parse tree generates a large number of hypothesis functions, thus the cost of parsing is amortized across many hypothesis functions.

This form of encoding can be used for other parse structures such as entity-relationship extraction.

\stitle{Annotations: }
Existing machine models are trained from  massive corpus of manually annotated data.  This ranges from bounding boxes of objects in images to multi-word annotations for information extraction models.  Each annotation type is akin to a node type in a parse tree and can be transformed into a hypothesis function that emits $1$ when the annotation is present and $0$ otherwise.

Similarly, image datasets (e.g., Coco~\cite{lin2014microsoft}, ImageNet~\cite{deng2009imagenet}) contain annotations in the form of bounding boxes or individual pixel labels.  Both can be modeled as hypotheses functions that map a sequence of image pixels to a Boolean sequence of whether the pixel is labeled with a specific annotation.

\stitle{Finite State Machines: }
Regular expressions, simple rules, and pattern detectors are easily expressed as finite state machines that explicitly encode state logic.   Since each input symbol triggers a state transition, an FSM can be wrapped into a hypothesis function that emits the current state label after reading the symbol.  Similarly, the state labels can be hot-one encoded, so that each state corresponds to a separate hypothesis function that emits $1$ when the FSM is in the particular state, and $0$ otherwise.

\stitle{General Iterators:}  More generally, programs that can be modeled as iterative procedures over the input symbols can be featurized to understand if units are learning characteristics of the procedure.  As an example, a shift-reduce parser is a loop that, based on the next input character, decides whether to apply a production rule or read the next character:
{\small
\begin{verbatim}
    initialize stack
    until done
     if can_reduce using A->B     // reduce
       pop |B| items from stack
       push A
     else                         // shift
       push next char \end{verbatim}}
\noindent Any of the expressions executed, or the state of any variables, between each \texttt{push next char} statement that reads the next character, can be used to generate a label for the corresponding character.  For instance, a feature may label each character with the maximum size of the stack, or represent whether a particular rule was reduced after reading a character.

\subsection{Natively Supported Measures}
\label{sec:measures}
\sys supports two types of statistical measures.

\stitle{Independent Measures:} measure the affinity between a single unit and a hypothesis function and are commonly used in the RNN interpretation literature.  Examples in prior work include Pearson's correlation, mutual information~\cite{morcos2018importance}, difference of means, Jaccard coefficient~\cite{zhou2017interpreting}, all available in \sys by default. In general, \sys supports any UDF that takes two behavior vectors as input. Independent measures are amenable to parallelization across units, which \sys enables by default.

\stitle{Joint Measures:} compute the affinity between a group of units $U$ and a hypothesis $h$, and  scores for each unit $u\in U$. For instance, when using logistic regression, we jointly compute one score for the whole group of units (e.g., prediction accuracy), and we assign individual scores based on model's coefficients. The current implementation supports convex prediction models that implement incremental \texttt{train} and \texttt{predict} methods. By default, we the use the logistic regression with L1 regularization trained with SGD (we use the optimizer Adam and Keras' default hyper-parameters), and we report the F1 on 5-fold cross-validation. \sys also supports arbitrary Keras and ScikitLearn models, as well as a multivariate implementation of mutual information.

\subsection{Verification}\label{ss:verification}
We note that DNI is fundamentally a data mining procedure that computes a large number of pairwise statistical measures between many groups of units and hypotheses.  When looking for high-scoring units, the decision is susceptible to multiple hypothesis testing issues and can lead to false positives.    Most current DNI analysis either do not perform verification (e.g., are best effort), or use one of a variety of methods.  One method is to ablate the model~\cite{morcos2018importance,karpathy2015visualizing} (``remove'' the high scoring units) and measure its effects on the model's output. Although a complete treatment to address this problem is beyond the scope of this paper, \sys implements a perturbation-based verification procedure to ensure that the set of high scoring units indeed have higher affinity to the hypothesis function.  To do so, the procedure is akin to randomized control trials, where, for a given input record, we perturb it in a way to swap a single symbol's hypothesis behavior, and measure the difference in activations.

Formally, let $h()$ be a hypothesis function that has high affinity to a set of units $U$.  It generates a sequence of behaviors when evaluated over a sequence of symbols:
{\small$$h([s_1,\dots,s_{k-1},s_k]) = [b_1,\dots,b_{k-1},b_k]$$}
\noindent After fixing the prefix $s_1,\dots,s_{k-1}$, we want to change the $k^{th}$ symbol in two ways.  We swap it with a {\it baseline} symbol $s^b_k$ so that $b^b_k$ remains the same, and with a {\it treatment} symbol $s^t_k$ so that $b^t_k$ changes.
{\small\begin{align*}
h([s_1,\dots,s^b_k]) &= [b_1,\dots,b^b_k]\  s.t. \ b^b_k = b_k, s^b_k \ne s_k\\
h([s_1,\dots,s^t_k]) &= [b_1,\dots,b^t_k]\  s.t. \ b^t_k \ne b_k, s^t_k \ne s_k
\end{align*}}
Let $act(s)$ be $U$'s activation for symbol $s$, $\Delta^b_k = act(s^b_k)-act(s_k)$ be the change  activation for a baseline perturbation, and $\Delta^t_k$ be the change for a treatment perturbation.  Then the null hypothesis is that $\Delta^t_k$ and $\Delta^b_k$, across different prefixes and perturbations, are drawn from the same distribution.

For example, consider the input sentence ``He watched Rick and Morty.'', where the hypothesis function detects coordinating conjunctions (words such as ``and'', ''or'', ``but'').  We then perturb the input words in two ways. The first is consistent with the hypothesis behavior for the symbol ``\texttt{and}'', by replacing ``\texttt{and}'' with another conjunction such as ``\texttt{or}''.   The second is inconsistent with the hypothesis behavior, such as replacing ``\texttt{and}'' with ``\texttt{chicken}''.  We expect that the change in activation of the high scoring units for the replaced symbol (e.g., ``\texttt{and}'') is higher when making inconsistent than when making consistent changes.
To quantify this, we label the activations by the consistency of the perturbation and then measure the Silhouette Score~\cite{rousseeuw1987silhouettes}, which scores the difference between the within- and between-cluster distances.

Our verification technique is based on analyzing the effects of input perturbations on unit activations, however there are a number of other possible verification techniques.  For instance, by perturbing the model using ablation~\cite{morcos2018importance,karpathy2015visualizing} (removing the high scoring units and retraining the model) and measuring its effects on the model's output.  We leave an exploration of these extensions to future work.

\section{System Design}
\label{s:opts}

\sys is implemented in Python and Keras, however it is also possible to embed DNI analysis into an ML-in-DB system such as MADLib~\cite{hellerstein2012madlib} through judicious use of UDFs and driver code.  This section describes two baseline designs---MADLib-based design and the naive \sys design---and their drawbacks.  It then introduces pragmatic optimizations to accelerate \sys.

\subsection{Baseline Designs}

\subsubsection{DB-oriented Design}
\label{sec:naive}

Using a database can help manage the massive unit and hypothesis behavior matrices that can easily exceed the main memory~\cite{vartak2016m}.  Also, as discussed in Section~\ref{sec:api}, it can be easier for users to post-process DNI results with relational operators (filtering, grouping, joining).   We now describe our DB-oriented implementation that uses the MADLib~\cite{hellerstein2012madlib} PostgreSQL extensions to perform DNI.

ML-in-database systems~\cite{hellerstein2012madlib,kumar2013hazy,feng2012towards} such as MADLib express and execute convex optimization problems (e.g., model training) as user-defined aggregates.  The following query trains a SVM model over records in \texttt{data(X, Y)} and inserts the resulting model parameters in the \texttt{modelname} table.
{\small\begin{center}
  \texttt{SELECT SVMTrain(`modelname', `data', `X', `Y'); }
\end{center}}
\noindent Note that the relation names are parameters, and the UDA internally scans and manipulates the relations.
An external process still needs to extract unit and hypothesis behaviors from the test dataset and materialize them as the relations \texttt{unitsb} and \texttt{hyposb}, respectively. Their schemas \texttt{(id, unitid/hypoid, symbolid, behavior)} contain the behavior value for each unit (or hypothesis) and input symbol.  This can be quite expensive.  After loading, a Python driver then submits one or more large SQL aggregation queries to compute the affinity scores.  For example, the correlation between each unit and hypothesis can be expressed as:
{\small\begin{verbatim}
   SELECT U.uid, H.h, corr(U.val, H.val)
     FROM unitsb U, hyposb H GROUP BY U.uid, H.h
\end{verbatim}}
The first challenge is behavior representation.    Deep learning frameworks~\cite{chollet2015keras,abadi2016tensorflow,paszke2017pytorch} return behaviors in a dense format.  Reshaping the matrices into a sparse format is expensive, and this representation is inefficient because it needs to store a hypothesis or unit identifier for each symbol. To avoid this cost, we can store the matrices in a dense representation where each unit (\texttt{U.uid\_i}) or hypothesis (\texttt{H.h\_j}) is an attribute. We compute the metrics as follows:
{\small\begin{verbatim}
 SELECT corr(U.uid1, H.h_1),...corr(U.uidn, H.h_m)
   FROM unitsb_dense U JOIN hyposb_dense H ON
        U.symbolid = H.symbolid
\end{verbatim} }
\noindent Unfortunately, there can easily be $>100k$ pairs of units/hypotheses to evaluate, while existing databases typically limit the number of expressions in a clause to e.g., 1,600 in PostgreSQL by default. We could  batch the scores (i.e., the sub-expressions \texttt{corr(U.uid, H.h\_m)}) into smaller groups and run one \texttt{SELECT} statement for each batch, but this would force PosgesSQL to perform hundreds of passes over the behavior relations (one full scan for each query). The problem is even more acute with MADLib's complex user-defined functions, such as \texttt{SVMTrain}, which incurs a full scan of the behavior tables and a full execution of the UDF for every hypothesis (see \Cref{s:exp_scale}). This leads to our second challenge: how to efficiently evaluate hundreds, potentially thousands of units/hypotheses pairs without incurring duplicate work?

The third challenge is that extracting the behavior matrices can be expensive~\cite{vartak2018m}. Unit behaviors require running and logging model behaviors for each record, while hypothesis behaviors require running potentially expensive UDFs.  For instance, our experiments use NLTK~\cite{nltk} for text parsing, which is slow and ultimately accounts for a substantial portion of execution costs. Furthermore, users often only want to identify high affinity scores, thus the majority of costs may compute low scores that will eventually be filtered out. Thus, it is important to reduce: the number of records that must be read, the number of unit behaviors to extract and materialize, the number of hypotheses that must be evaluated, and affinity score computation that are filtered out.

Our experiments find that this baseline is far slower than all versions of \sys, and point to the bottlenecks to address in order to support deep neural inspection within a database system.


\subsubsection{Naive \sys Design}

\begin{figure}[tb]
  \centering
  \includegraphics[width=\columnwidth]{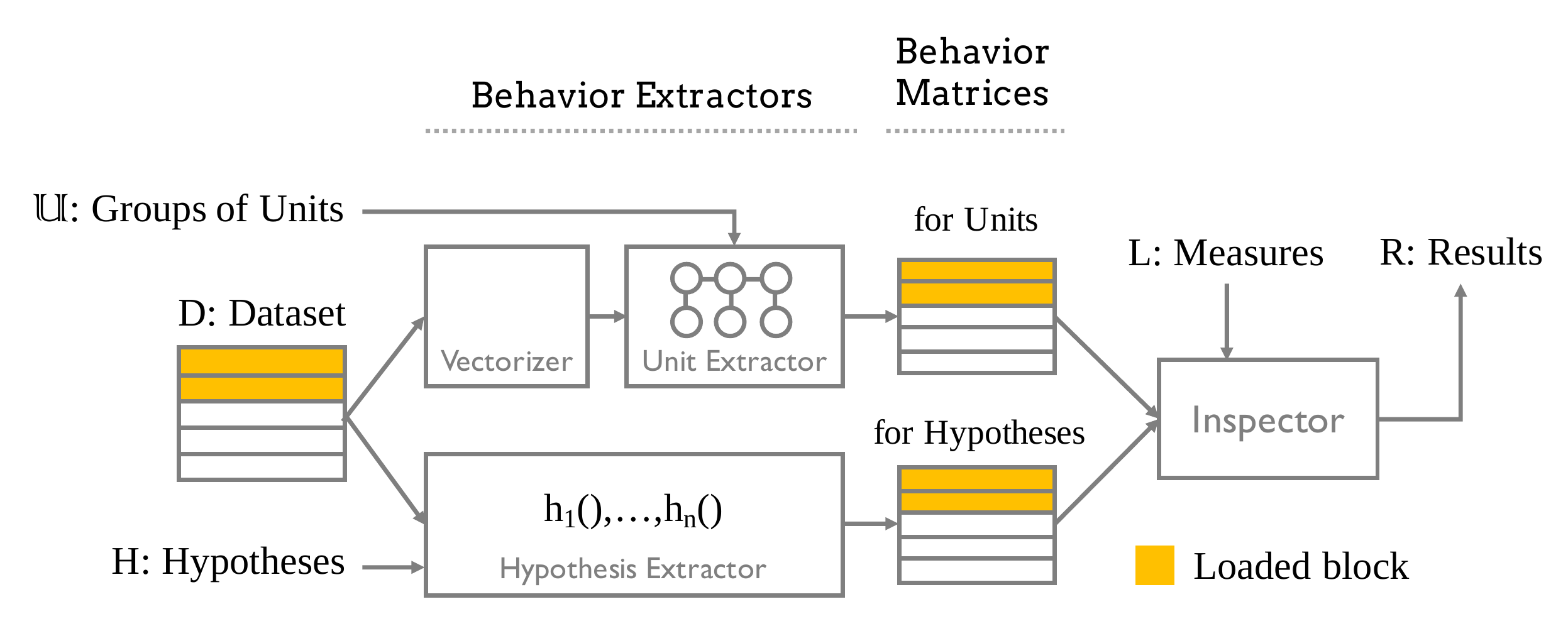}
  \caption{\sys Architecture.}
  \label{f:arch}
  \vspace*{-.2in}
\end{figure}

\Cref{f:arch} presents the naive \sys architecture.  Its major drawback is the need to excessively materialize intermediate matrices.  The design will be optimized in the next subsection.

\sys first materializes all behaviors from the dataset $D$. The {\it Unit Behavior Extractor} takes one or more unit groups as input, and generates behaviors for each unit in each group---the assumption is that each group is a subset of units from a single model.  Similarly, the {\it Hypothesis Behavior Extractor} takes a set of hypothesis functions as input and runs them to generate hypothesis behaviors.  We concatenate the sequences together, so the extractors emit matrices of dimensionality $ |D|\cdot n_s \times |\mathbb{U}| $ (for Units) and $ |D|\cdot n_s \times |H|$ (for Hypotheses).  Since the number and length of records ($|D|\cdot n_s$) can dwarf the number of units and hypotheses, these matrices are ``skinny and tall''.  The \texttt{Inspector} takes as input these two matrices and desired statistical measures, and computes affinity scores for each unit, hypothesis, and measure triplet.

\sys natively supports activation extraction for Keras models, but can be extended with custom Unit Extractors for other frameworks such as PyTorch, or to simply read behaviors from pre-extracted files. For instance, our experiments in Section~\ref{sec:nmt} use on a custom PyTorch extractor for the OpenNMT model. Any object that inherits our class \texttt{Extractor} and that exposes the following method may be used in \sys:
\begin{center}
  \texttt{extract(model, records, hid\_units) $\rightarrow$ behaviors}
\end{center}
\noindent The input \texttt{records} is a list of records, and \texttt{hid\_units} is a list of integers that uniquely identify hidden units. The output \texttt{behaviors} is a NumPy array with one row per symbol and one column per hidden unit. The user may pass additional arguments (e.g., batch size) in the constructor of the class. Note that this is the minimal API, a few additional methods must be written to support the optimizations presented in following sections.

\sys{} extracts unit activations using a GPU, which accelerates activation extraction as compared to a single CPU core.  Hypotheses are executed using a single CPU core.    \sys trains logistic regression models, and more generally all affinity measures based on linear models, as a Keras neural network model on a GPU.    Finally, \sys can cache the hypothesis behavior matrix in cases where the model repeatedly changes. Our implementation uses simple LRU to pin the matrix in memory, and integrating caching systems such as Mistique~\cite{vartak2018m} for unit and hypothesis behaviors is a direction for future work.

\subsection{Optimizations}
\label{optimizations}
Below we outline the main optimizations.

\subsubsection{Shared Computation via Model Merging}
Although affinity score measures are typically implemented as Python User Defined Aggregates, \sys also supports Keras computation graphs. For instance, the default Logistic Regression measure is implemented as a Keras model. This enables a shared computation optimization we call model merging.  The naive approach trains a separate model for every hypothesis, which can be extremely expensive. Instead, \sys \emph{merges} the computation graphs of all $|H|$ hypotheses into a single large composite model.  The composite model has one output for each hypothesis rather than $|H|$ models with one output each.  This lets \sys make better use of Keras' GPU parallel execution.  It also amortizes the per-tuple overheads across the hypotheses---such as shuffling and scanning the behavior relations, and data transfer to the GPU. This optimization is exact, it does not impact the final scores.

\sys produces one composite model for each affinity measure.  For a given measure's Keras model, it duplicates the intermediate and final layers for each hypothesis and enforces them to share the same input layers.  Thus they share the input layer, but maintain separate outputs.  If the model doesn't have a hidden layer (as in logistic regression), \sys can further merge all output layers into a single layer with one or more units (if the categorical output is hot-one encoded) per hypothesis; \sys then generates a global loss function that averages the losses for each hypothesis. This optimization does not degrade the results: since there is no dependency between the models and their parameters, minimizing the sum of the losses is equivalent to minimizing each loss separately. We do however lose the ability to early-stop the training for the individual hypotheses, as we cannot freeze individual hidden units in Keras.

Model merging is applicable when the \emph{scoring function} provided by the user is based on Keras (e.g., the logistic regression score in our experiments). This is orthogonal to the framework of the model to inspect---the optimization could very well support custom extractors for other frameworks, e.g., PyTorch.

\subsubsection{Early Stopping}
\label{sec:early-stop}
Much of machine learning theory assumes that datasets used to train machine learning models are samples from the ``true distribution'' that the model is attempting to approximate~\cite{efron2016computer}.  \sys assumes that the dataset $D$ is a further sub-sample.  Thus, the affinity scores are actually empirical estimates based on sample $D$.

A natural optimization is to allow the user to directly specify stopping criteria to describe when the scores have sufficiently converged.  To do so, a statistical measure $l()$ can expose an incremental computation API:
\begin{center}
  \texttt{l.process\_block(U, h, recs)$\rightarrow$(scores, err)}
\end{center}
\noindent The API takes an iterator over records \texttt{recs} as input and returns both the group and unit scores in \texttt{scores}, as well as an error of the group score \texttt{err}.  Users can thus specify a maximum threshold for \texttt{err}. If this API is supported, then \sys can terminate computation for the pair of units and hypothesis function early.  Otherwise, \sys ignores the threshold and computes the measure over all of $D$.

We expose an API rather than make formal error guarantees because such guarantees may not available for all statistical measures.  For example, tight error bounds are not well understood for training non-convex models (e.g., neural nets), and so in practice machine learning practitioners check if the performance of their model converges with empirical methods (i.e., comparing the last score to the overage over a training window~\cite{prechelt1998early}). There exists however formal error bounds for statistical measures such as correlation~\cite{efron2016computer}.

By default \sys implements this API for pairwise correlation and logistic regression models. To estimate error of the correlation score, we use Normal-based confidence intervals from the statistical literature (i.e., Fisher transformation~\cite{efron2016computer}). For logistic regression, we follow established model training procedures and report the difference between the model's current validation score and the average scores over the last $N$ batches, with $N$ set up by default to cover 2,048 tuples.

Early stopping is implemented by iteratively loading and processing blocks of pre-materialized unit and hypothesis behavior matrices in blocks of $n_b$ records.  Records on disk are assume to have been shuffled record-wise.   It loads for all units in each group $U$, and for as many hypotheses as will fit into memory, and checks the error for every statistical measure after each block. We shuffle the blocks symbol-wise in-memory before running inspection. The SGD based approaches shuffle the behaviors further during training.

Note that the moment the score for a given hypothesis and unit group has converged, then there is no need to continue reading additional blocks for that hypothesis.  Thus, there is a natural trade-off between processing very small blocks of rows, which incurs a the overhead of checking convergence  more frequently, and large blocks of rows, which may process more behaviors than are needed to converge to $\epsilon$.  Empirically, we find that setting $n_b = 512$ works well because most measures converge within a few thousand records.  This optimization ensures that the query latency is bound by the complexity of the statistical measure, rather than the size of the test dataset.

\subsubsection{Streaming Behavior Extraction}
A consequence of employing approximation is that \sys does not need to read all of the materialized matrices.  Our third optimization is to materialize the behavior and hypothesis matrices in an online fashion, so that the amount of test data that is read is bound by how quickly the confidence of the statistical measures converge.   To do so, we read input records in blocks of $n_b$ and extract unit and hypothesis behaviors from them in parallel.  An additional benefit of this approach is that affinity scores can be computed and updated progressively, similar to online aggregation queries, so that the user can stop \sys after any block.

Figure~\ref{f:arch} illustrates streaming execution using the orange blocks.  The input $D$ contains $5$ blocks of records, and only two blocks of unit and hypothesis behaviors have been extracted so far.  When all affinity scores have converged, then \sys can stop.  Although it is possible to further optimize by terminating hypothesis extraction for hypotheses that have converged, we find that the gains are negligible.  This is because 1) training the composite model from model merging costs the same for one hypothesis as it does for all, and 2)  some hypothesis extractors, such as creating a parse tree for NLP, incurs a single cost amortized across all parsing-based hypotheses derived from the parse tree.

\section{Experiments}
\label{s:experiments}

Our experiments study the scalability of \sys, as well as its ability to generate DNI scores that are comparable to prior DNI analyses.   To this end, we first  present scalability experiments using a SQL auto-completion RNN model to show how the baselines, \sys, and its optimizations scale as we vary the number of hidden units, hypotheses, and records.   We then use \sys to analyze a real world English-to-German translation model from OpenNMT~\cite{opennmt}, and report results from reproducing DNI analysis from Belinkov et al~\cite{belinkov2017neural}.

We provide additional experiments in the Appendix. In \Cref{s:accuracy}, we present a set of experiments to evaluate the accuracy of \sys's scores. In \Cref{a:exp}, we complete our SQL auto-completion scalability benchmark by commenting the results provided by the system. In \Cref{a:dni-cnn}, we extend our analysis to convolutional neural nets and compare \sys to NetDissect, an existing system to analyze computer vision models.

\subsection{Setup Overview}
We ran \sys on two types of RNN models: the first predicts the next symbol (character) for SQL query strings generated from a subset of the SQL grammar, while the second is a sequence-to-sequence English to German translation model called OpenNMT.

\stitle{Datasets:}  We used two language datasets: a collection of SQL queries for the scalability benchmark, and a publicly available English-to-German translation dataset for the real-world experiment\footnote{http://statmt.org/wmt15}. To generate synthetic SQL queries, we sample from a Probabilistic Context Free Grammar (PCFG) of SQL.  We choose subsets of the grammar (between 95 to 171 production rules) to vary the language complexity, as well as the number of hypothesis functions.  The task is to take a window of $30$ characters and predict the character that follows.

\stitle{Models:} The SQL use-case is based on custom models: a one-hot encoded input layer, a LSTM layer, and a fully connected layer with soft-max loss for final predictions (details below).  The OpenNMT model~\cite{opennmt} is publicly available, it uses an encoder-decoder architecture, where both the encoder and decoder contain two LSTM layers of 500 units, with an additional attention module for the decoder.

\stitle{Hypotheses: }
For the SQL experiment, we follow the procedure in Section~\ref{ss:hypotheses} to transform parse trees into a set of hypothesis functions.  By default, we use the time-domain representations for each node type (e.g., production rule, verb, punctuation).  In our experiments, we do not run the parser until one of the hypothesis functions is evaluated; at that point the other hypothesis functions based on the parser do not need to re-parse the input text. To increase the number of hypotheses, we also generate hypothesis functions using the signal representation.  We use NLTK's chart parser~\cite{nltk} to sample and parse the SQL grammar. For the translation experiments, we use Part-of-Speech tagger of CoreNLP\cite{manning2014corenlp}, which we can directly use as a hypothesis.

\subsection{Scalability Benchmarks}\label{s:exp_scale}
We now report scalability results on the SQL grammar benchmark.  To do so, we vary the number of records in the inspection dataset, hidden units in the model and rules in the grammar used to generate the data and the features. The default setup contains 29,696 records\footnote{Recall that each record in \sys is a window of symbols of length $n_s$ as defined by a sliding window of size $n_s$ and stride 5.}, 512 hidden units and 142 grammar rules. Each record has $n_s=30$ symbols, so there are 890,880 behaviors for each unit and hypothesis.

We build two hypotheses per non-terminal in the grammar.  The first one returns ``1'' for each symbol for which the rule is active (the symbol is consumed by the rule or a descendant rule).  The second only returns ``1'' for the first and last symbol and returns ``0'' otherwise.  This yields 190 hypotheses.

\stitle{Systems: } We start with the Python baseline implementation \pybase, then cumulatively add the optimizations described in Section~\ref{s:opts}: model merging (\naive), early stopping (\appr), and online extraction (\sys).  In addition, we measure the benefits of the GPU by comparing the model-merging baseline with a GPU (\gpu) and without (\cpu).   We compare against the \madlib implementation, which fully materializes the behavior matrices, and then computes affinity scores using PostgreSQL native (for correlation) and MADLib (for logistic regression) functions.

We run each configuration 3 times report and the average. For each experiment, we run the smallest-scale baseline to completion, and then enforce a 30-minute timeout for larger-scale settings.

\stitle{Setup:} All our experiments are based on 6 Google Cloud virtual machines with 32GB RAM running Ubuntu 16.04, and 8 virtual CPUs each, where each virtual CPU is a hyper-thread on a 2.3 GHz Intel Xeon E5 CPU.  Each VM includes a nVidia Tesla K80 GPUs with 12 GB GDDR5 memory. All models are based on Keras with Tensorflow 1.8. MADLib uses PostgreSQL 9.6.9, with the shared buffer size, effective cache size, and number of workers tuned following to the manual's guidelines. Hypothesis extraction is performed by creating a parse tree using NLTK~\cite{nltk} and transforming the tree into many hypotheses.

We extract behaviors in blocks of 512 records, and set the Keras batch size to 512 records.  All the models are trained for up to 50 epochs with Keras early stopping. Their average classification accuracy is $49.7\%$,  and 53-69\% of randomly generated queries can be parsed (based on the grammar complexity).
The approximation defaults use $\epsilon=0.025$ and 95\% confidence for correlation, and error threshold of $0.01$ for logistic regression (cf. Section~\ref{sec:early-stop}).

\begin{figure}[h]
  \centering
  \includegraphics[width = \columnwidth]{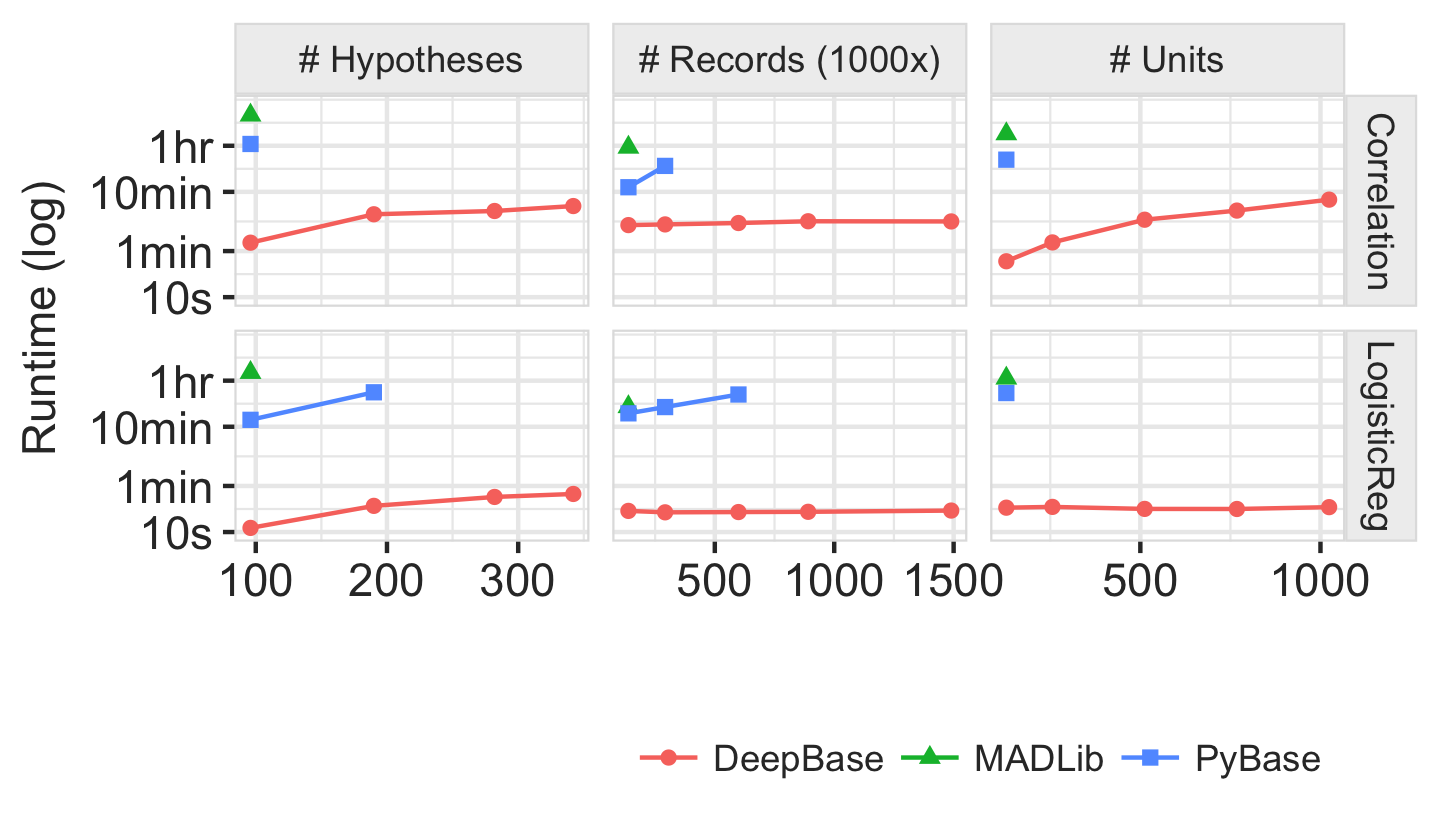}
  \caption{Runtime of MADLib and Python baselines as compared to \sys with all optimizations for logistic regression measure.}
  \label{f:exp-baselines}
\end{figure}
\stitle{Comparing Baselines:}
Figure~\ref{f:exp-baselines} compares the MADLib and Python baseline systems for both affinity measures (rows) as we vary the number of hypotheses, records, and hidden units in the model (columns).  We also include \sys with all optimizations for reference.

Correlation (top row) is generally expensive because it must be computed for every unit-hypothesis pair (up to 194,560 pairs in the experiments). \madlib incurs a large number of passes over the behavior relations (up to 121).  Both \madlib and \pybase  incur considerable full table scan and aggregation costs.
Logistic regression (bottom row) is dominated by the cost to fit logistic regression models for each pair of hypothesis and unit group.  

Overall, we find that \pybase performs faster than \madlib on the smallest experimental settings.  We believe this is largely because of the overheads of using PostgreSQL extensions and the fact that the in-memory Python implementation of logistic regression is quite fast.
\sys's optimizations avoids unnecessary extraction costs once all the scores have converged.
\sys improves upon \pybase by $72\times$ on average, and by up to $96\times$; it outperforms \madlib by $200\times$ on average, and by up to $419\times$.

\begin{figure}[h]
  \centering
  \includegraphics[width = \columnwidth]{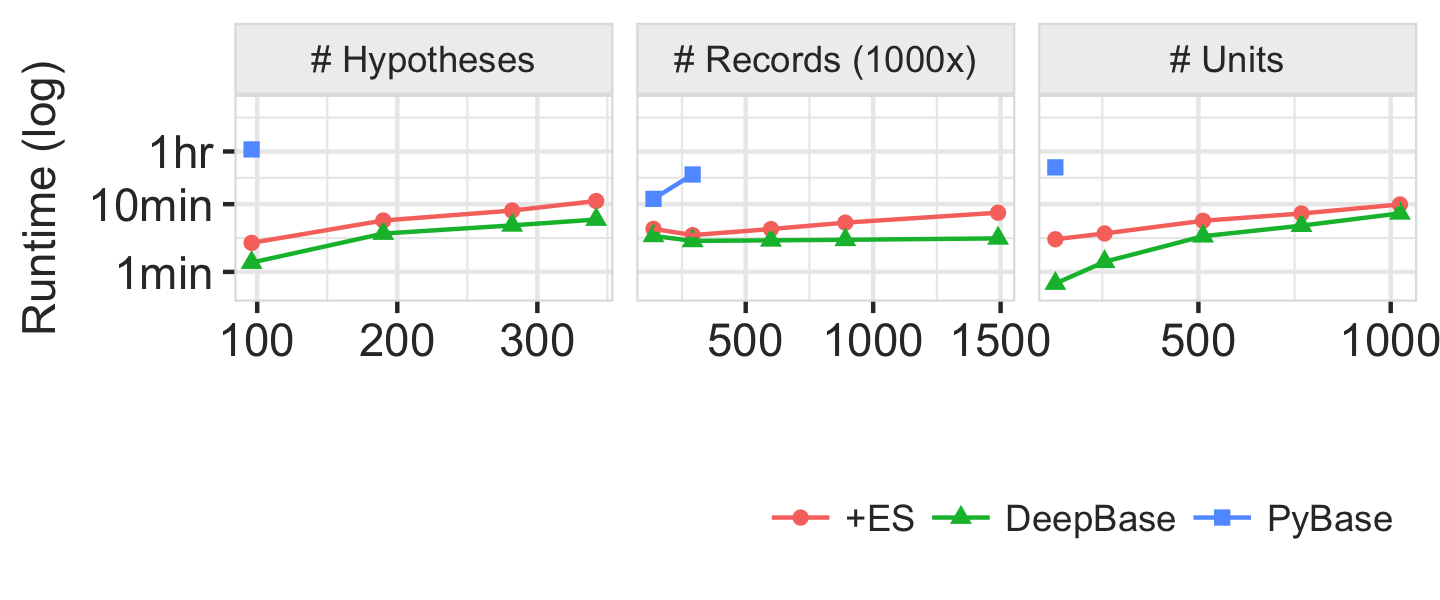}
  \vspace*{-.4in}
  \caption{Runtime of \sys with different optimizations enabled for correlation measure.}
  \label{f:exp-scale-opt-corr}
\end{figure}

\stitle{Optimization Benefits for Correlation Measure:}
Figure~\ref{f:exp-scale-opt-corr} reports runtimes for three variants of \sys for correlation.  Correlation is a cheap measure to compute and is executed on the CPU.  Since we use a CPU, model merging (which is an GPU-oriented optimization) is disabled.  Thus we compare \pybase, with early stopping, and with lazy extraction.

We find that the primary performance gains are due to the early stopping optimization, while lazily extracting behaviors provides considerable, but smaller benefit.  We see that lazy extraction provides a benefit as the number of records increases (middle plot), and similarly, the benefit of lazy extraction reduces as the number of hidden units increases (right plot) because the bottleneck becomes the large number of pair-wise correlation computations.

\begin{figure}[h]
  \centering
  \includegraphics[width = \columnwidth]{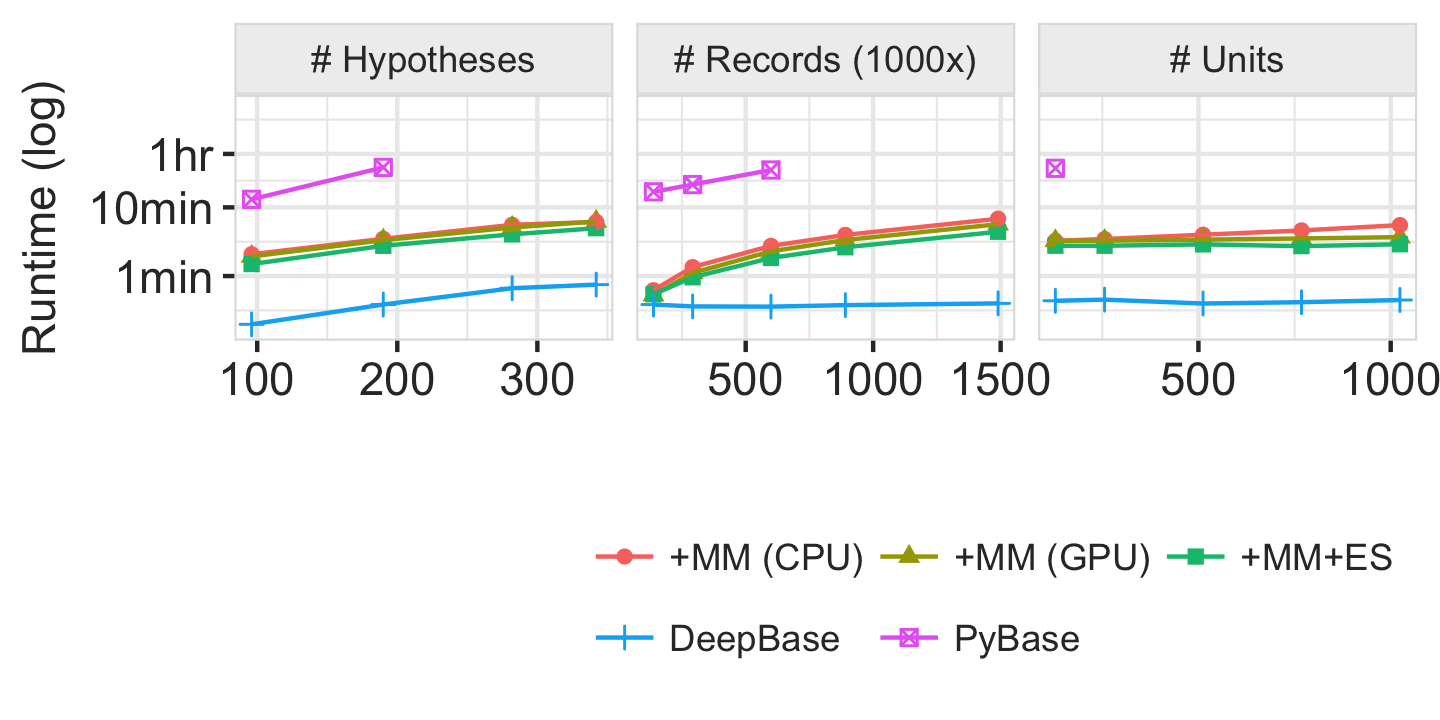}
  \vspace*{-.4in}
  \caption{Runtime of \sys with different optimizations enabled for logistic regression measure.}
  \label{f:exp-scale-opt-log}
\end{figure}

\stitle{Optimization Benefits for Logistic Regression Measure:}
Figure~\ref{f:exp-scale-opt-log} reports the results when adding optimizations based on early stopping, lazy extraction as well as GPU-based optimizations.    We see that model merging (\naive) provides a considerable benefit by reducing the number of logistic regression models that need to be trained for each hypothesis.  The benefits of using a GPU appear for models with many hidden units.  We find that early stopping (\appr) does not provide any speedup because materializing the behavior matrices is a large bottleneck; adding lazy extraction (\sys) reduces the runtime by $6\times$ on average and by up to $11\times$ as compared to \appr.

\begin{figure}[h!]
  \centering
  \includegraphics[width =.8 \columnwidth]{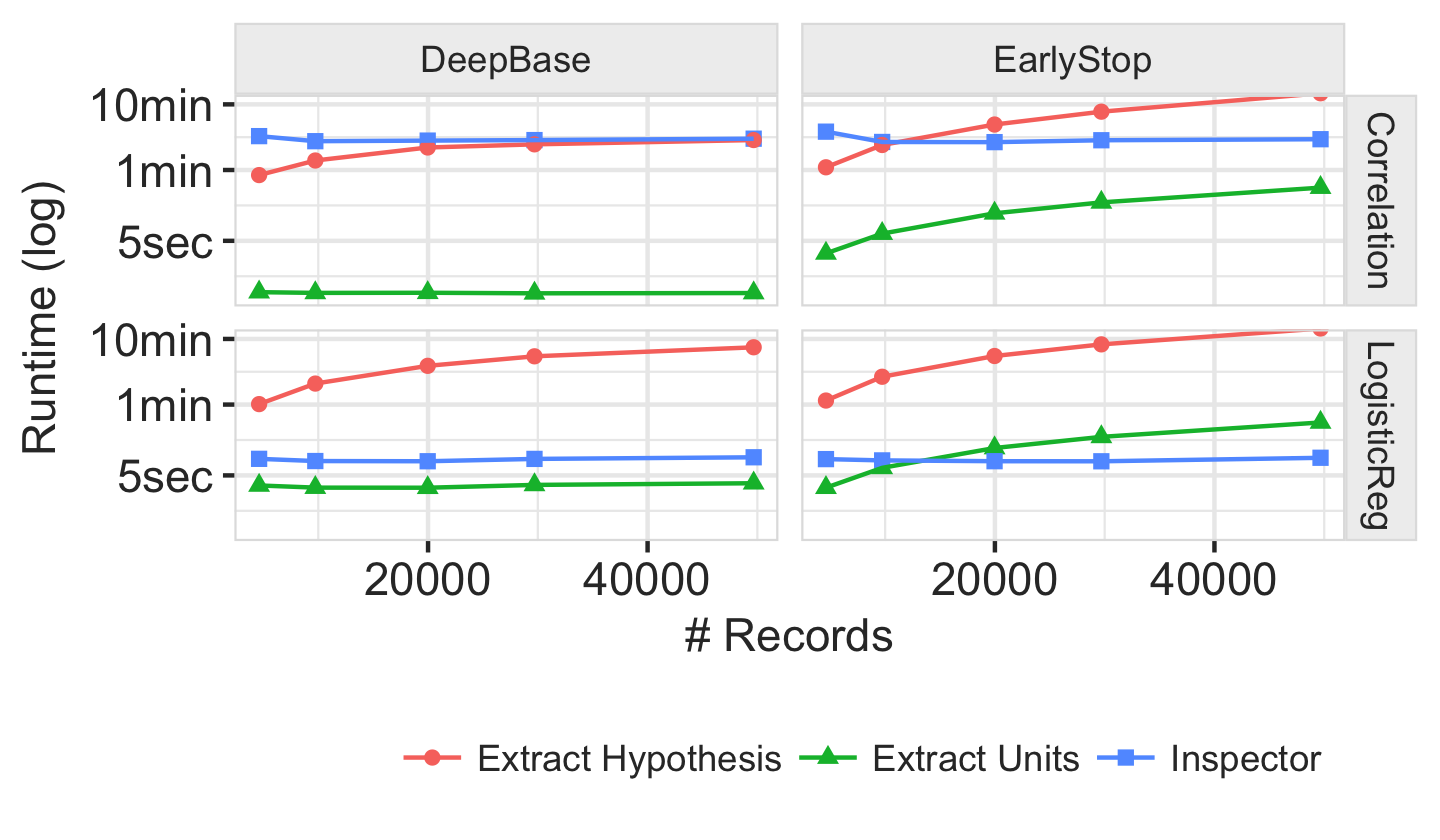}
  \vspace*{-.2in}
  \caption{Runtime breakdown of extraction and inspector costs for correlation and logistic regression.}
  \label{f:exp-scale-breakdown}
\end{figure}

\stitle{Runtime Breakdown: }  Figure~\ref{f:exp-scale-breakdown} shows the cost breakdown by system component: the hypothesis and unit extractors, and the inspector.  The \appr column  shows that inspector cost is much higher for correlation, while extraction behave nearly identically.  The \sys column shows that runtime savings are primarily due to lower extraction costs thanks to online extraction.

\begin{figure}[h!]
  \centering
  \includegraphics[width = .9\columnwidth]{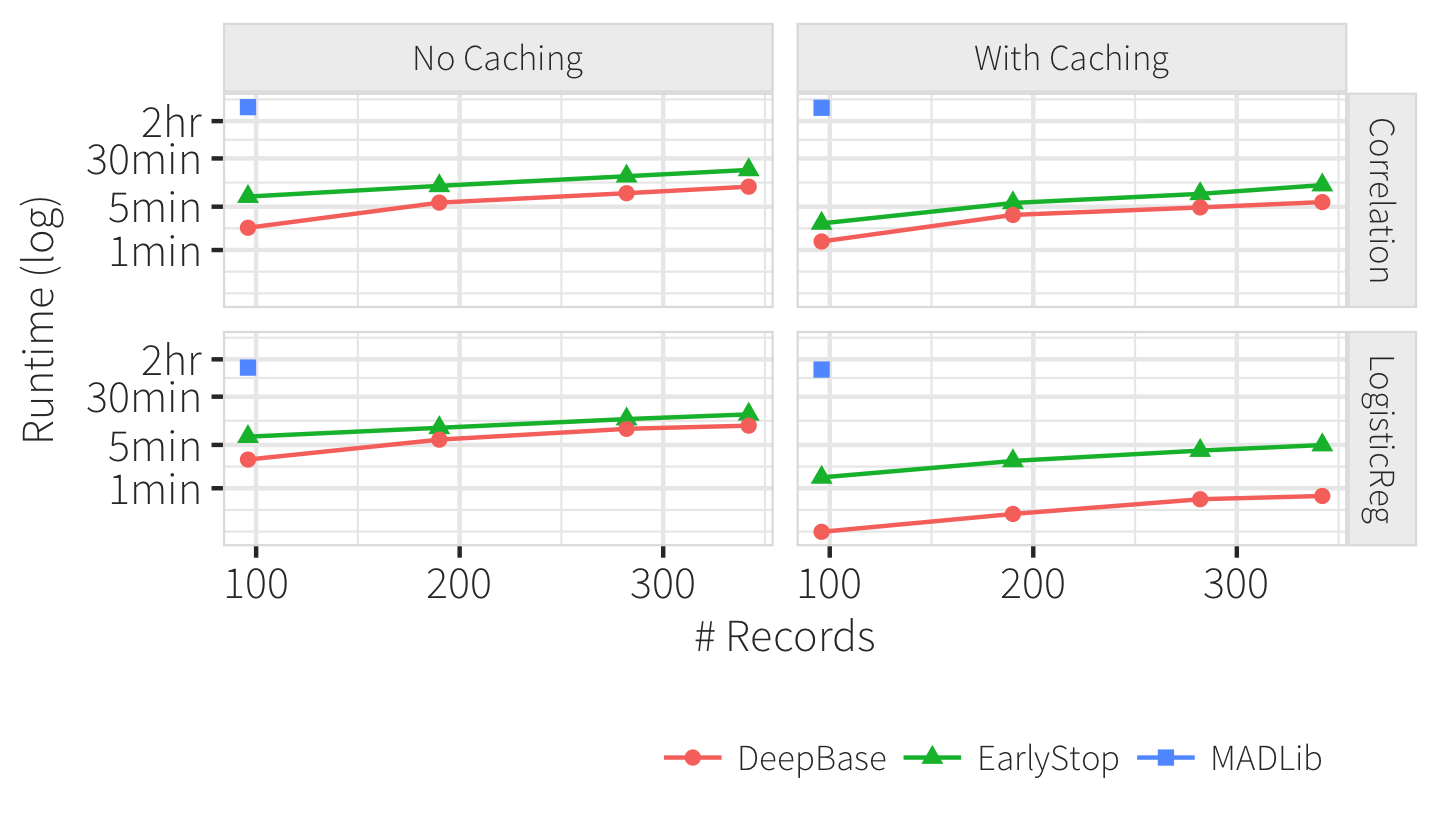}
  \vspace*{-.2in}
  \caption{Runtime comparing the effects of cached hypothesis behavior.}
  \label{f:exp-scale-runtime-noparse}
\end{figure}

\stitle{Cached Hypothesis Extraction: }
We found that hypothesis extraction due to a slow parsing library can dominate the runtime.   However, during model development or retraining, the developer typically has a fixed library of interesting hypothesis functions and wants to continuously inspect how the model behavior is changing.  \Cref{f:exp-scale-runtime-noparse} examines this case: the left column incurs all runtime costs, while the right column shows when hypothesis behavior has been cached.  We see that it improves correlation somewhat, but its cost is dominated by inspection; whereas for logistic regression, \sys converges to $\approx20s$.  Caching improves correlation by $1.9\times$ on average, and logistic regression by $12.4\times$ on average and up to $19.5\times$.  Overall, \sys outperforms \madlib by up to $413\times$.

\begin{figure}[h!]
  \centering
  \includegraphics[width = .9\columnwidth]{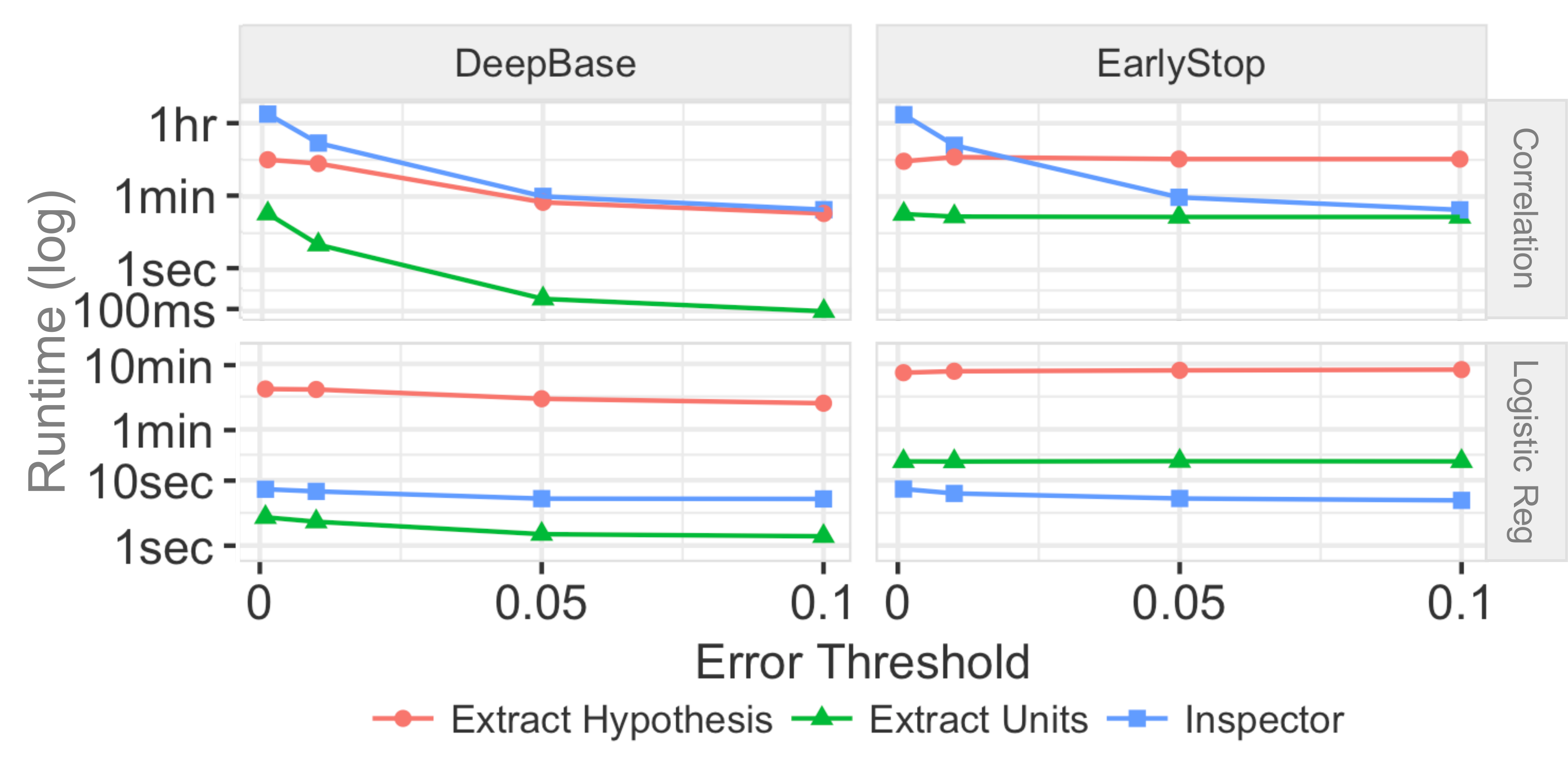}
  \vspace*{-.2in}
  \caption{Runtime when varying error threshold for early stopping.  Note different Y-axis scales.}
  \label{f:exp-scale-error}
\end{figure}

\stitle{Sensitivity to Error Threshold: }
Figure~\ref{f:exp-scale-error} examines the sensitivity to varying the error threshold (x-axis) for correlation and logistic regression, using the default experiment parameters.  The top row shows correlation: \appr only reduces the inspector costs as the threshold is relaxed, while \sys reduces the extraction costs considerably because it only extracts behaviors when necessary.  The bottom row shows logistic regression, which exhibits similar trends, though it is far less sensitive to the error threshold because the optimization converges slowly.

{\it \stitle{Takeaways: }
Independent measures such as correlation are computed on a per-unit basis, and the cost is dominated by inspection costs.  In contrast, joint measures such as logistic regression are computed for each unit-group are dominated by extraction costs.  Early stopping improves independent measure performance, while online extraction enables \sys to run nearly independently of dataset size.  \sys outperforms the MADLib and Python baselines by two orders of magnitude and took at most $10.3$ minutes for the slowest setting.
}

\subsection{Neural Machine Translation}
\label{sec:nmt}

We reproduce the analysis of existing studies by applying \sys on an public English-German translation model. We first replicate the methodology of Belinkov et al.~\cite{belinkov2017neural} and verify that our results are consistent with those returned by their scripts\footnote{\url{github.com/boknilev/nmt-repr-analysis}}. We then broaden the analysis and show that we can make observations that are similar in spirit to those presented in related work~\cite{shi2016emnlp, adebayo2018local} with only a few queries.


In addition to those results, we present a comparison of \sys with NetDissect~\cite{bau2017network}, a recent interpretation method for convolutional Neural Nets in Appendix~\ref{a:dni-cnn}.

\subsubsection{Comparison with Belinkov et al.}

Belinkov et al. have shown that sequence-to-sequence neural translation models learn part-of-speech tags as a byproduct of translation. They train a classifier from the encoder's hidden layer activations and observe that they can predict the tags of the input words with high accuracy. This section replicates this analysis.

We ran \sys and the baseline scripts on the same datasets, using the same 46 POS tags and the same score function. We used an English-to-German translation corpus available online\footnote{\url{drive.google.com/file/d/0B6N7tANPyVeBWE9WazRYaUd2QTg/view}}, annotated with the Stanford CoreNLP tagger\footnote{\url{stanfordnlp.github.io/}}. We use 4,823 sentences for training, 636 for validation and 544 for testing (each sentence contains 24.2 words in average). Our score is the precision of a multi-class logistic regression model trained on the encoder's hidden unit outputs, as described in the the original paper. We limit to 35 training epochs, with a patience (i.e., number of epochs without improvement before early-stopping) of 5, the scripts' default.

\stitle{Experimental Setup Differences:} An important difference between the two approaches is that they run in different environments. Belinkov et al. supports \texttt{seq2seq-attn} models, a legacy library that runs on top of Torch for Lua. Unfortunately, we found no way to import \texttt{seq2seq-attn} models into Python. After consulting the library's authors, we chose to support \texttt{seq2seq-attn}'s successor, \texttt{OpenNMT}, which runs in PyTorch for Python. We extended \sys extraction library to support this model. Because of this incompatibility, we run each system on a different model. For \sys, we use a public model from OpenNMT, a sequence-to-sequence model with 2 LSTM layers of 500 hidden units each, available online\footnote{\url{opennmt.net/}}. For Belinkov et al., we use a custom model trained with \texttt{seq2seq-attn} and strived to replicate OpenNMT's setup as closely as possible: we used the same NN architecture and training data, and similar training parameters. We expect the results of the analysis to be strongly correlated, but not identical, because the training and experiments environment is not identical, and the two models are implemented differently.

\begin{figure}[t!]
  \centering
  \includegraphics[width=0.4\columnwidth]{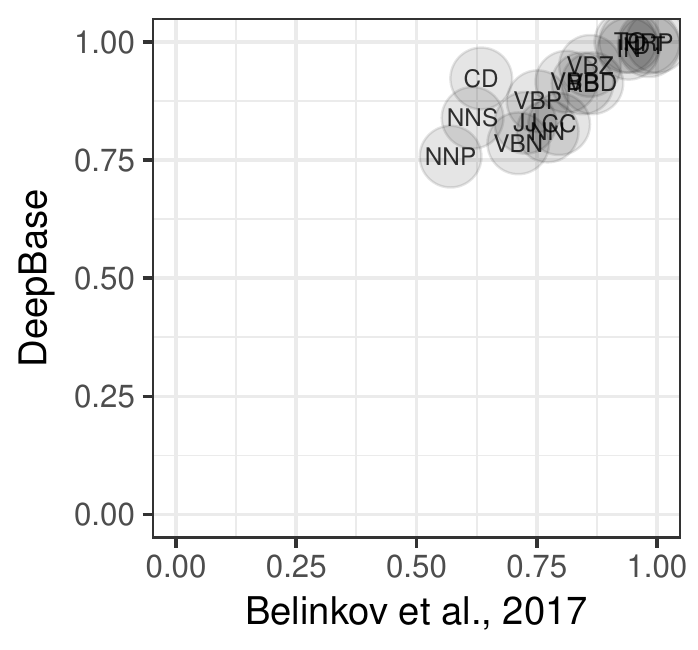}
  \caption{Precision for every tag, computed by Belinkov et al. and by \sys. We filtered out the tags that cover less than 1.5\% of the data. The sample Pearson correlation is r=0.84.}
  \label{f:bel-vs-dni}
\end{figure}

\stitle{Results:} Figure~\ref{f:bel-vs-dni} presents the affinity scores for every POS tag computed by the two approaches. The strong correlation (0.84) between the approaches suggest analysis consistency.

Belinkov et al.'s scripts ran in 1 hour and 10 minutes, while \sys ran in 55.1  minutes. Aside from the differences in frameworks, we explain the difference as follows. Belinkov et al. modify the NMT model in-place by freezing the weights of the translation model and inserting the POS tag classifier directly in the encoder. Since the dataset is relatively small, they must make many passes over the data before the classifier converges (at least 35), running the full translation model each time. By contrast, \sys extracts the activations once (this takes 38.3 minutes) and makes the subsequent passes on the cached version (7.4 minutes), which amortizes the activation extraction time. Note that none of \sys's optimizations apply to this use case: model merging is irrelevant because there is only one hypothesis (the function is not binary, it returns one the 42 distinct POS speeches at each step), and early stopping/lazy materialization does not help because the dataset is small.

{\it \stitle{Takeaways: } \sys can easily express the analysis presented \cite{belinkov2017neural}, its scores are consistent with the scripts provided by the authors and its runtime is competitive.}

\subsubsection{Additional Results}
\begin{figure}[th!]
  \centering
  \begin{subfigure}[t]{.4\textwidth}
  \centering
  \includegraphics[width = \columnwidth]{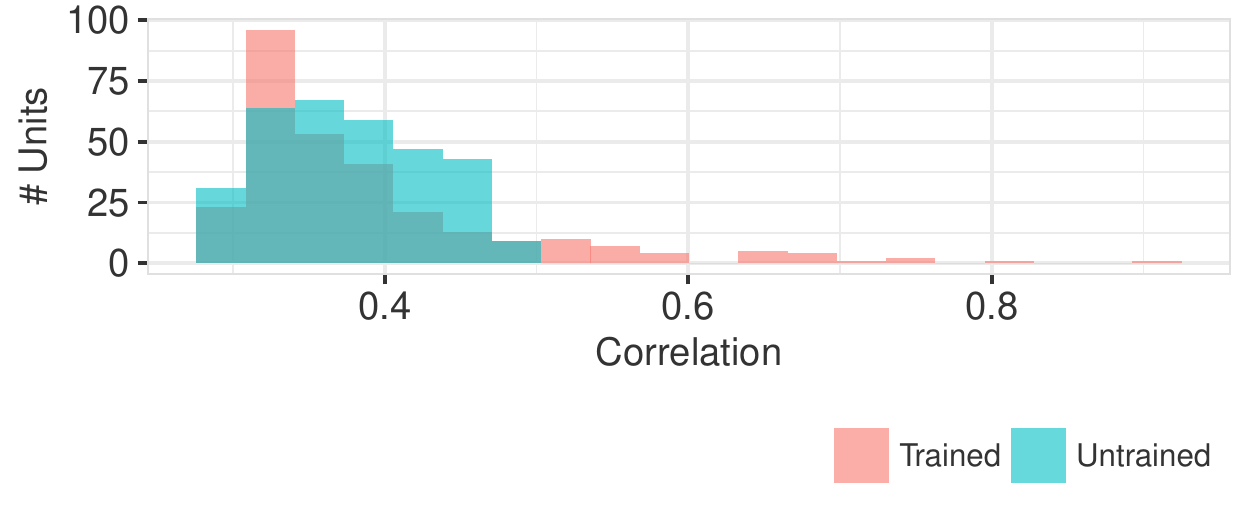}
  \caption{\small Histogram of correlations for all encoder units in OpenNMT.  High correlations are only found in the trained model.}
  \label{f:exp-nmt-corr}
  \end{subfigure}

   \begin{subfigure}[t]{.4\textwidth}
  \centering
  \includegraphics[width = \columnwidth]{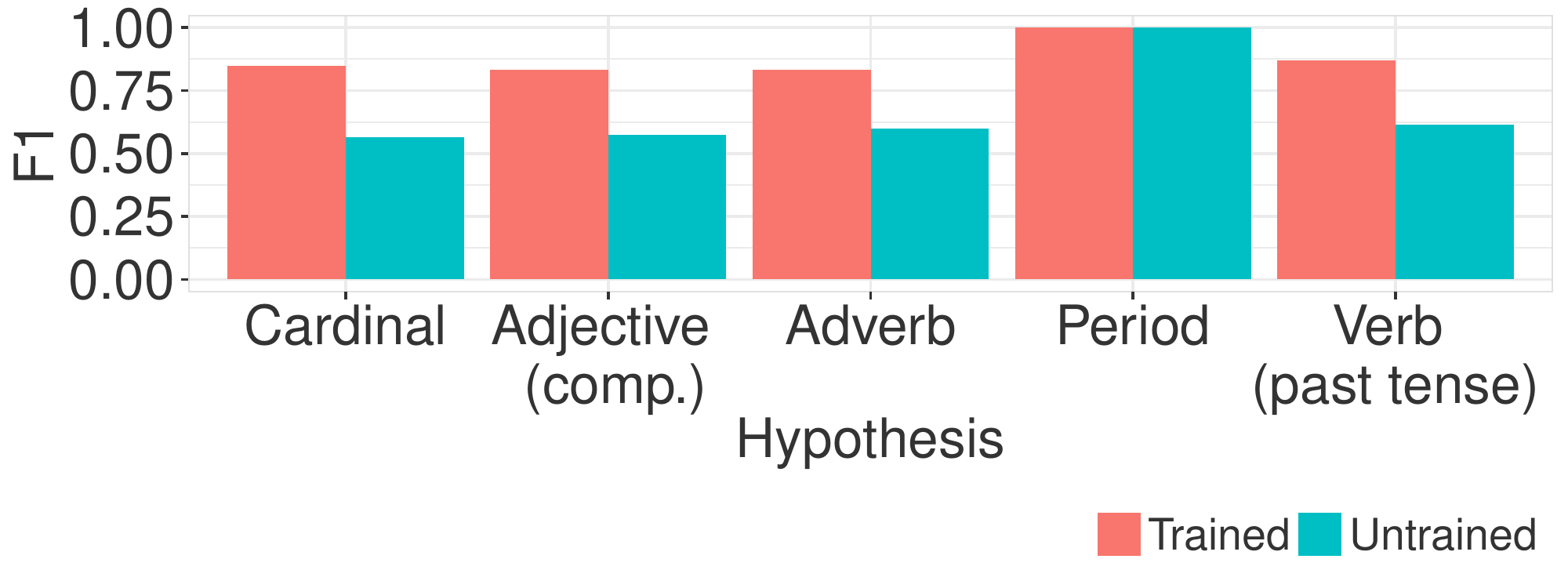}
  \caption{\small L2 Logistic Regression F1 measure for different hypotheses.  Both  models learn low-level hypotheses (period); only the trained model learns higher level concepts.}
  \label{f:exp-nmt-logreg}
  \end{subfigure}
  \vspace*{-.1in}
  \caption{\small Deep neural inspection on OpenNMT translation model. Results compared against an untrained OpenNMT model.  }
  \label{f:exp-nmt}
\end{figure}
We now broaden our analysis, and show that we can replicate and verify the conclusion of recent work~\cite{karpathy2015visualizing, adebayo2018local, shi2016emnlp} with only a few additional queries. For the remainder of those experiments, we add $7$ hypotheses for phrase-level structures (NP, VP, PP, etc.) to the POS presented previously.

\stitle{Individual Units: } We first use correlation to study individual units. Prior work found individual interpretable units in character-level language models~\cite{karpathy2015visualizing}, and we find similar units at the word-level. They learn low-level features (e.g., periods, commas, etc) along with one unit that tracks the sentence length.  Going beyond past studies, we find that high affinity units are only present in the trained model and not in an untrained model of the same architecture (Figure~\ref{f:exp-nmt-corr}).

\stitle{Encoder Level: } We then use logistic regression with L2 regularization to study all $1000$ units in a trained and untrained model (Figure~\ref{f:exp-nmt-logreg}).  We first confirm recent work showing that model architecture can act as a strong prior~\cite{adebayo2018local}.  Similarly, the untrained model has high affinity with some low level language features (e.g, periods), but low affinity for almost all high-level features.  On the other hand, the trained model has
far higher affinity to various POS tags (e.g., CD, RB, VBD, etc.) and phrase structure (e.g., VP, NP) than the untrained model.

\stitle{Unit groups: }   We now inspect each layer separately, and use Logistic Regression with L1 to identify unit groups with non-zero coefficients.  Previous work~\cite{belinkov2017neural,shi2016emnlp} showed that both encoder layers learn POS features, but layer $0$ is slightly more predictive and more distributed (spread over more units).  Similarly, we find that layer $0$ yields higher F1 scores and selects more units for most hypotheses. Going beyond prior work, we find that the unit group size varies widely depending on the language feature. In layer $1$ for example, $372$ units are found to detect verbs, $62$ units to detect coordinating conjunctions (e.g. `and', `or', `but'), while only $9$ units to detect punctuation such as ``.''.

{\it \stitle{Takeaways: }\sys expresses and computes results that are consistent with analyses in recent NLP studies that seek to understand neural activations in machine translation~\cite{belinkov2017neural,shi2016emnlp,shi2016emnlp}, with orders of magnitude less engineering effort.  Our library of natural language hypothesis functions automates model inspection for syntactic features that NLP researchers are commonly interested in, and is easy to extend.  Our declarative API lets users easily inspect and perform different analyses by comparing different models at the granularity of individual, groups of, and layers of hidden units.}
\section{Related Work}
\label{sec:related}

\stitle{Interpreting Neural Networks:} Many approaches were proposed for model interpretation. Section~\ref{s:background} reported three
methods: visualization of the hidden unit
activations~\cite{hermans2013training, karpathy2015visualizing,
strobelt2018lstmvis}, saliency analysis~\cite{qiqi17icml, girshick2014rich,
li2015visualizing, simonyan2013deep, selvaraju2016grad, zhou2014object} and statistical neural inspection~\cite{noroozi2016unsupervised, alain2016understanding, kim2017tcav, kadar2017representation, bau2017network, morcos2018importance}. These methods are common in the neural net understanding literature, and motivate the design of \sys. Other approaches generate synthetic inputs by inverting the transformation induced by the hidden layers of a neural net (the most compelling example reveal e.g., textures, body parts or objects)~\cite{nguyen2016synthesizing,nguyen2016plug}. However, most of the literature focuses on computer vision, and the process relies heavily on human inspection. Another form of analysis is occlusion analysis, by which machine learning engineers selectively replace patches of an image by a black area and observe which hidden units are
affected as a result~\cite{zeiler2014visualizing}. Currently, most studies that fall under
this category are ad-hoc and target image analysis. Our verification method~\ref{ss:verification} is an attempt to generalize and automate this process by defining input perturbations (of which occlusion is one type of input perturbation) with respect to the desired hypothesis function.

Because the field is still in its infancy~\cite{doshi2017towards, hohman2018visual}, the majority of existing implementations are specialized research prototypes and there is a need for general software systems in the same way TensorFlow and Keras simplify model construction and training.   A notable example is Lucid~\cite{olah2018the},  which bundles feature inversion, saliency analysis, with visualization into a larger grammar.  Lucid has similar goals as \sys, however it focuses on images and still relies on manual analysis.

\stitle{Visual Neural Network Tools: }
Numerous visualization tools have been developed to inspect the architecture of deep models~\cite{kahng2018cti, smilkov2017direct}, do step debugging to check the validity of the computations~\cite{cai2016tensorflow}, visualize the convergence of gradient descent during training~\cite{kahng2016visual} and drill into test sets to understand where models make errors~\cite{kahng2016visual}.

\stitle{Machine Learning Interpretation: }
A related field of research seeks to augment machine learning predictions with explanations, to help debugging or  augment software produces based on classifier. A common, classifier-oblivious approach is surrogate models, which approximates a complex model by a simpler one (e.g., classification tree or logistic regression).  They train a simple, often linear, model over examples in the neighborhood of a test data point so that users can interpret the rationale for a specific model decision~\cite{ribeiro2016should, alvarez2017causal}.   Other approaches modify the machine learning model so that predictions are inherently interpretable, such as PALM~\cite{krishnan2017palm}. In contrast, \sys seeks to identify general behaviors with respect to a test dataset by inspecting individual and groups of unit behaviors.

\stitle{Databases and Models: }
A number of database projects have proposed integrating machine learning models, training, and prediction into the database~\cite{boehm2014hybrid,kumar2017data,feng2012towards,deshpande2006mauvedb,hellerstein2012madlib}.  Recent projects such as ModelDB~\cite{vartak2016m} and Modelhub~\cite{miao2016modelhub} propose to manage historically trained models and can be used by \sys to select models and hidden units to inspect.  Similarly, systems such as Mistique~\cite{vartak2018m} can be used in conjunction with \sys to manage the process of extracting and caching unit activations.

\section{Conclusion and Discussion}

Programming frameworks for deep learning have enabled machine learning to impact a large set of applications.  Yet efforts to understand and inspect the behaviors of hidden units in NN models are largely manual and one-of, and require considerable expertise and engineering effort.   Better NN analysis tools will contribute to a better understanding of how and why neural networks work. Towards this goal, this paper defined {\it Deep Neural Inspection} to  characterize existing inspection analyses, and presented \sys, a system to quickly inspect neural network behavior through a declarative API.  With a few lines of code, \sys can express a large fraction of existing deep neural inspection analyses, but improves the analysis run-times by up to $72\times$ as compared to existing baseline designs.  Further, we reproduced results consistent with prior NLP research~\cite{belinkov2017neural} on real-world translation models~\cite{opennmt}.

We intend to extend \sys with more statistical measures and deeper integration with GPUs, support distributed environments, and apply \sys to a broader range of applications (e.g., bias detection, reinforcement learning). Looking further, we envision Deep Neural Inspection as a core primitive of a larger neural network verification and inspection framework~\cite{sellamlike}.  \sys allows users, or automated processes to query neural network models using high level hypotheses.  We imagine curating {\it libraries of hypotheses} based on decades of existing models, features, and annotations across application domains.  In addition, these tools may be used to decompose NNs into smaller components, enforce activation behavior for unit groups, and ultimately open up NN black boxes.

\section{Acknowledgments}
We acknowledge: NVIDIA Corporation for donating a nVidia Titan XP GPU, the Google Faculty Research Award, the Amazon Research Award, and NSF grants 1527765 and 1564049. Thanks to Yonathan Belinkov for his code, guidance, and advice~\cite{belinkov2017neural}, and Yiliang Shi for advice.
\balance

\clearpage

{\small
\bibliographystyle{abbrv}
\bibliography{main}
}

\appendix
\section{Neural Network Primer}\label{sec:primer}
We provide a brief review of neural networks. For an extensive overview, we refer readers to Goodfellow et al.\ \cite{Goodfellow-et-al-2016}.

Neural networks are mappings that transform an input vector $s$ to an output vector $y$. The mapping is parameterized by a vector of weights $w$, which is learned from data. To capture nonlinear relationships between $s$ and $y$, we can stack linear transformations followed by nonlinear functions:
\begin{align}
y &= w_{n+1}^T h_n \\
h_n &= \sigma\left(w_n^T h_{n-1}\right) \\
h_0 &= s
\end{align}
where $h_n$ is a vector of \emph{hidden units} in the $n$th layer that represent intermediate states in the neural network, which are also frequently called activations or ``neurons''.  The function $\sigma$ must be nonlinear for the model to learn nonlinear mappings, and today most neural networks use the rectified linear unit $\sigma(x) = \textrm{max}(0, x)$. When interpreting neural networks, we are often interested in understanding what the hidden units $h_n$ are learning to detect, which is the focus of this paper.

Recurrent neural networks (RNNs) are popular models for operating on sequences, for example processing text as a sequence of words. Given an input sequence where $s_t$ is the $t$-th element in the sequence, a recurrent network follows the recurrence relation:
\begin{align}
y_t &= w_2^T h_t \\
h_t &= \sigma\left( w_1^T \left[s_t, h_{i-1}\right]\right) \\
h_0 &= 0
\end{align}
where the intermediate hidden units $h_t$ are a function of the $t$th element in the sequence and the previous hidden units from the previous element in the sequence.  Importantly, the parameters $w$ are independent of the position in the sequence, allowing the model to process arbitrary length sequences. Modern recurrent networks use a more sophisticated update mechanism to stabilize optimization, for example with LSTMs \cite{hochreiter1997long}, which we use in our experiments. However, the high-level question of interpretation remains the same: we are interested in analyzing and understanding the intermediate activations $h_i$ in the network.

Recall that $h_t$ is a vector with multiple dimensions. For clarity, we refer to a specific dimension in $h_t$ as $u_{i}$, which represents a single hidden unit.
We call any value that can be computed for $u_i$ at each time step a {\it behavior}. For example,
we can compute measures such as $u_i$'s gradient with regards to the input (the derivative $\frac{\delta u_i}{\delta s_t}$).

\section{SQL Extensions via \texttt{INSPECT}}\label{ss:inspect}
This section describes how DNI can be integrated into a SQL-like language as a new \texttt{INSPECT} clause.  We introduce a separate clause because DNI is neither a scalar UDF nor an user defined aggregation (UDA). Instead, it outputs a set of records for each input group (UDAs return a single record per group), and the \texttt{INSPECT} operator needs to flatten the groups into a single relation before sending to other relational operators.

\stitle{NNs as Relations:} \sys models hidden units, hypotheses, and model inputs as relations (or views) in a database.  Let \texttt{units} be a relation representing hidden units (\texttt{uid}) and their models (\texttt{mid}), and \texttt{hypotheses} represent hypothesis functions (\texttt{h}). These relations may contain additional meta-data attributes---e.g., unit layer, training epoch, or the source of the hypothesis function---for filtering and grouping the hidden units.

\stitle{\texttt{INSPECT}:} The syntax specifies unit ids and hypothesis functions to compare, optional affinity measures (e.g., logistic regression), and the dataset of input sequences used to extract unit and hypothesis behaviors.  By default, \sys measures correlation between individual units and hypotheses:
{\small\begin{verbatim}
    INSPECT <unit>, <hypothesis> [USING metric, ...]
       OVER <sequences>
\end{verbatim}}
\noindent The clause is evaluated prior to the \texttt{SELECT} clause and outputs a temporary relation with schema \texttt{(uid, hid, mid, group\_score, unit\_score)} containing unit, hypothesis, and model ids, and two affinity scores.  \texttt{group\_score} is the affinity between a group of units $U$ and the hypothesis $h$, whereas \texttt{unit\_score} specifies the affinity of each unit $u\in U$; groups are defined using \texttt{GROUP BY}. These scores are interpreted depending on the type of statistical measure.  For instance, correlation is computed for individual units (each group is a single unit), so the two scores are the same.  In contrast, when using logistic regression, the model F1 score represents the affinity of the group, while the coefficients are the individual unit scores.
This relation can be renamed but only referenced in later clauses (e.g., \texttt{SELECT}, \texttt{HAVING}).

\stitle{SQL Integration: }
Users often inspect models as part of debugging.  We now show the full query syntax to express the DNI analysis from the motivating example in \Cref{ss:motivation}.
The query groups hidden units by the sql parser model's training epochs, computes the correlation between each unit in layer $0$ with a hypothesis that recognizes SQL keywords (e.g., ``SELECT'', ``FROM``), and returns the epoch and id of high scoring units:
{\small\begin{verbatim}
   SELECT M.epoch, S.uid
  INSPECT U.uid AND H.h USING corr OVER D.seq AS S
     FROM models M, units U, hypotheses H, inputs D
    WHERE M.mid = U.mid AND M.mid = 'sqlparser' AND
          U.layer = 0 AND H.name = 'keywords'
 GROUP BY M.epoch
   HAVING S.unit_score > 0.8
\end{verbatim}}
\noindent The user can easily change the layer or types of models to inspect with slight modifications of the query, or further join the output with other analysis queries:

\begin{figure*}[t!]
  \centering
   \begin{subfigure}[t]{.40\textwidth}
      \includegraphics[width = \columnwidth]{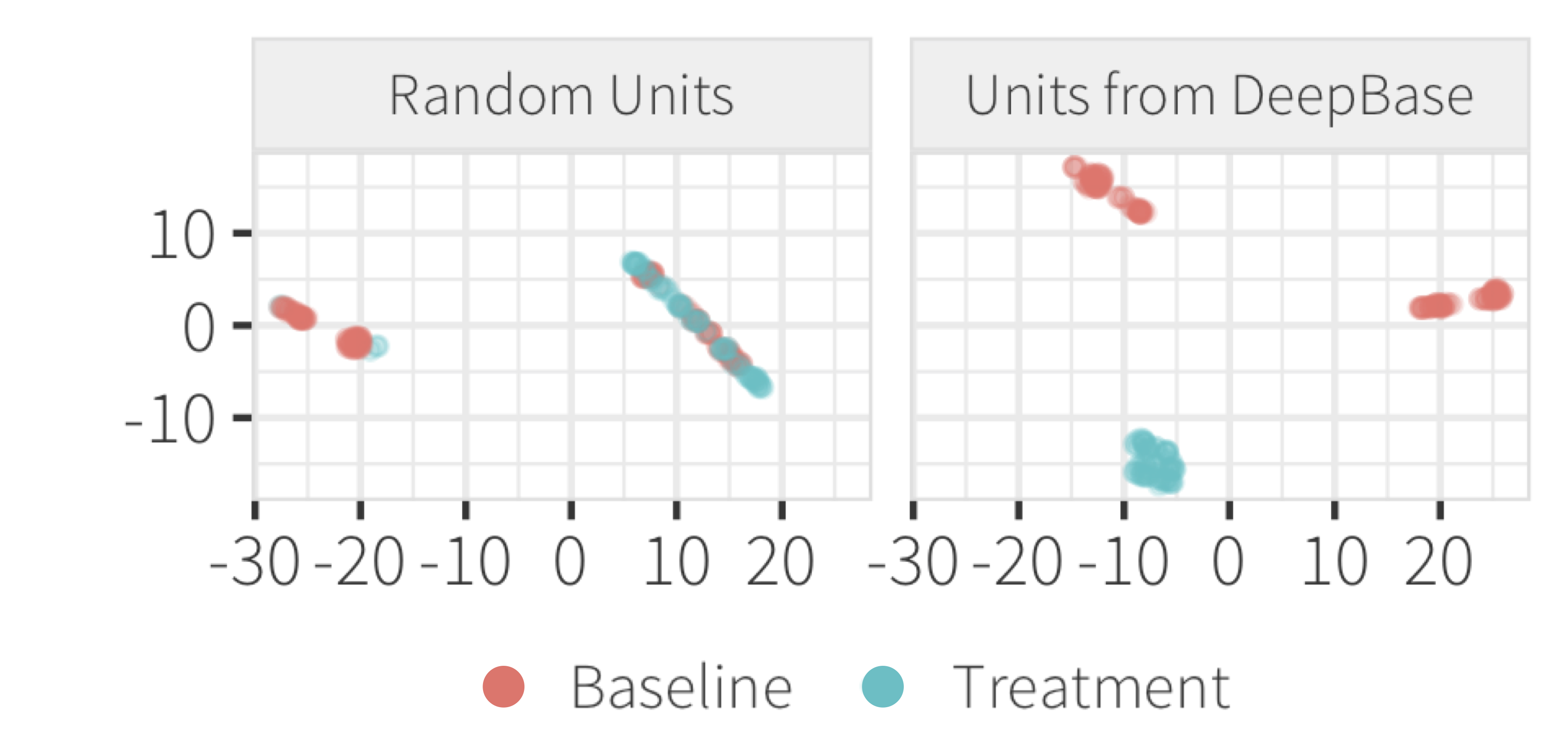}
      \vspace*{-.1in}
      \caption{T-SNE clustering of change in activation during verification. Each point is one unit's activation. \\ (Weight=0.5, Num Specialized=4)}
      \label{f:exp-acc-spec-tsne}
  \end{subfigure}
  \begin{subfigure}[t]{.29\textwidth}
    \centering
    \includegraphics[width = \columnwidth]{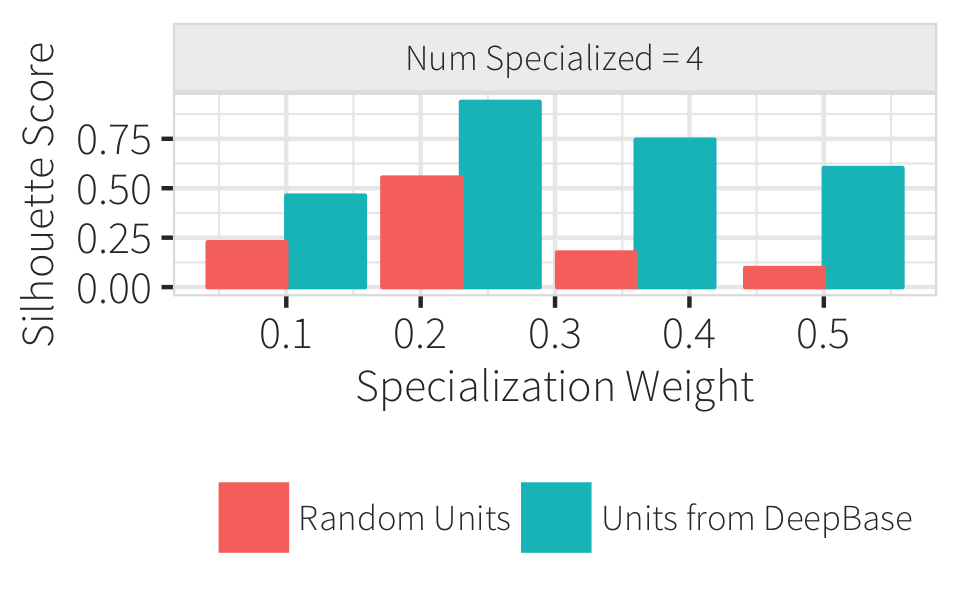}
    \vspace*{-.1in}
    \caption{Silhouette scores varying the number of specialized units. (Weight=0.5)}
    \label{f:exp-acc-spec-sil-group4}
  \end{subfigure}
  \begin{subfigure}[t]{.29\textwidth}
    \centering
    \includegraphics[width = \columnwidth]{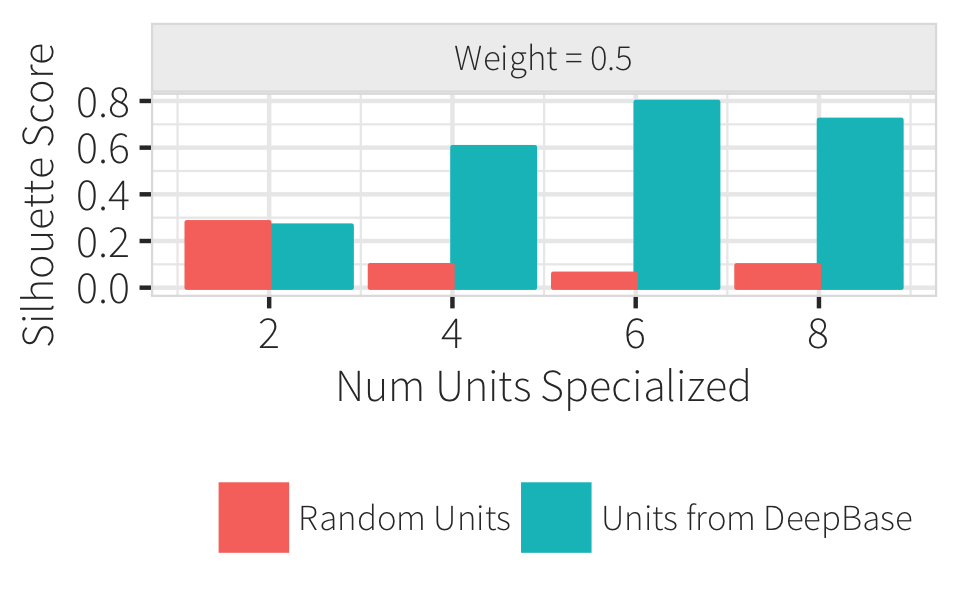}
    \vspace*{-.1in}
    \caption{Silhouette scores varying weight.  (Num Specialized=4)}
    \label{f:exp-acc-spec-sil-weight05}
  \end{subfigure}
  \vspace*{.1in}
  \caption{\small Verification results for parentheses detection hypothesis function.   }
  \label{f:exp-acc-spec}
\end{figure*}

\section{Accuracy Benchmark}\label{s:accuracy}
This set of experiments attempts to assess whether the hidden units that \sys scores highly are indeed correct.   To this end, we first present an accuracy benchmark to study DNI under conditions when we ``force'' parts of a model to learn a hypothesis function.

\stitle{Dataset:} We generated a dataset by sampling from a Probabilistic Context Free Grammar (PCFG). The dataset (used in~\cite{strobelt2016visual}) consists of strings such as \texttt{0(1(2((44))))} where a digit representing the current nesting level may precede each balanced parenthesis (up to 4 levels).  Its grammar consists of the same production rule
$r_i \rightarrow i \; r_i \, | \, \texttt{(}  r_{i+1} \texttt{)} \ $
for $i<4$ nesting levels, along with a terminating rule $r_4 \rightarrow \epsilon \, | \, \texttt{4} \; r_4$ for the $4^{th}$ level.

\stitle{Setup:} To establish ground truth, we specially train a 16-unit RNN model by ``specializing'' a subset of units  $S\subseteq M$ to learn a specific hypothesis $h$. To do so, we introduce an auxiliary loss function $g_h$ that forces the output of the neurons in $S$ to to be close to the output of $h$.  If $g_h$ is the auxiliary loss function and $g_T$ is the loss function for $M$ based on the next character prediction task, the model's loss function is a weighed average of $g_M = w\times g_h + (1-w)\times g_T$.  This setup allows us to vary the number of specialized units $|S|$, and the specialization weight $w$ for how much the specialized units focus on learning the hypothesis.  Their defaults are $|S|=4$, $w=0.5$.

The challenge in this benchmark is that not all units in $S$ may be needed to learn a given hypothesis.  For instance, if the hypothesis is to detect the current input, then one unit may be sufficient.  In contrast, all units may be needed to learn a higher level hypothesis.    We run \sys using logistic regression with L1 regularization, return units with unit\_score above $15$, and use the perturbation-based verification method in Section~\ref{ss:verification} to assess the quality of the high scoring units.

\stitle{Results:} Figure~\ref{f:exp-acc-spec-tsne} shows an example clustering of the change in activations between the baseline and treatment perturbations (colors).  The hypothesis function used to inspect the model recognizes parentheses symbols, thus the baseline perturbations swap `$($' for `$)$' or vice versa.   The treatment swaps `$($' for a number.  Units selected by \sys show clear clusters that distinguish baseline and treatment perturbations, while the change in activation for a set random units (the same number of units) overlap considerably (blue and red are indistinguishable). Figure~\ref{f:exp-acc-spec-sil-group4} summarizes the cluster separation using the Silhouette score~\cite{rousseeuw1987silhouettes} and shows higher separation than random units across all weights.  Similar results are shown when varying the number of specialized units in Figure~\ref{f:exp-acc-spec-sil-weight05}.

We also ran the above for two other hypothesis functions: predicting the current nesting level, and predicting that the current nesting level is 4.  The former hypothesis is nearly identical to the model task, and we indeed find that none of the units selected by \sys distinguish themselves from random during verification.  The latter hypothesis is ambiguous: the specialized units may simply recognize the input character $4$, or learn the nesting level.   After running verification by swapping $4$ with other numbers (baseline) or open parentheses (treatment), we find that the change in activations were indistinguishable.  Thus suggests that the specialized units learned to recognize the input $4$ rather than the logical nesting level.

{\it \stitle{Takeaways:}
Although ground truth does not exist for deep neural inspection analyses, unit specialization provides a (weak) form of ground-truth. In general, DNI analyses (using \sys or another system) is a form of data mining, and may misinterpret the behavior of hidden units (e.g., if the hypothesis is very similar to the model task or ambiguous). \sys's perturbation-based verification method helps us identify these false positives.
}

\section{SQL Auto-complete Inspection}\label{a:exp}
This section extends the scalability experiments in \Cref{s:exp_scale} with an analysis of the inspection results.  We can use \sys to study what the model learns through its training process by executing a query akin to the example in Section~\ref{ss:inspect}.  We train the SQL auto-completion model by performing several passes of gradient descent over the training data, called epochs. We repeat the process until the model's performance converges or starts to decrease (after 13 epochs in our case). We capture a snapshot of the model after random initialization (then the accuracy is 1.1\%), 1st epoch (41\% acc), and 4th epoch (45\% acc), and perform neural inspection to understand what the model learned.

Figure~\ref{f:sql_top} shows a few of the highest affinity hypotheses.  These hypotheses correspond to fundamental SQL clauses that should be learned in order to generate valid SQL, and the model appears to learn them (rather than arbitrary N-grams) even in the first training epoch. Further, the F1 is higher for detecting the string ``ORDER'', which we expect is needed to learn the ordering expression \texttt{ordering\_term}.

\begin{figure}[t]
  \centering
\includegraphics[width = \columnwidth]{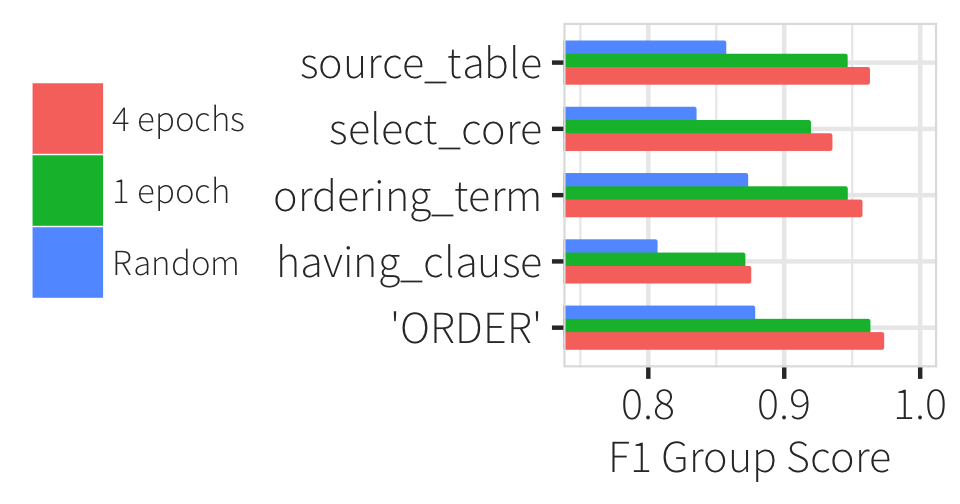}
  \caption{\small F1 scores for highest affinity hypotheses during training.}
  \label{f:sql_top}
\end{figure}

\section{\sys for CNNs}\label{a:dni-cnn}
\begin{figure}[b]
  \centering
\includegraphics[width = \columnwidth]{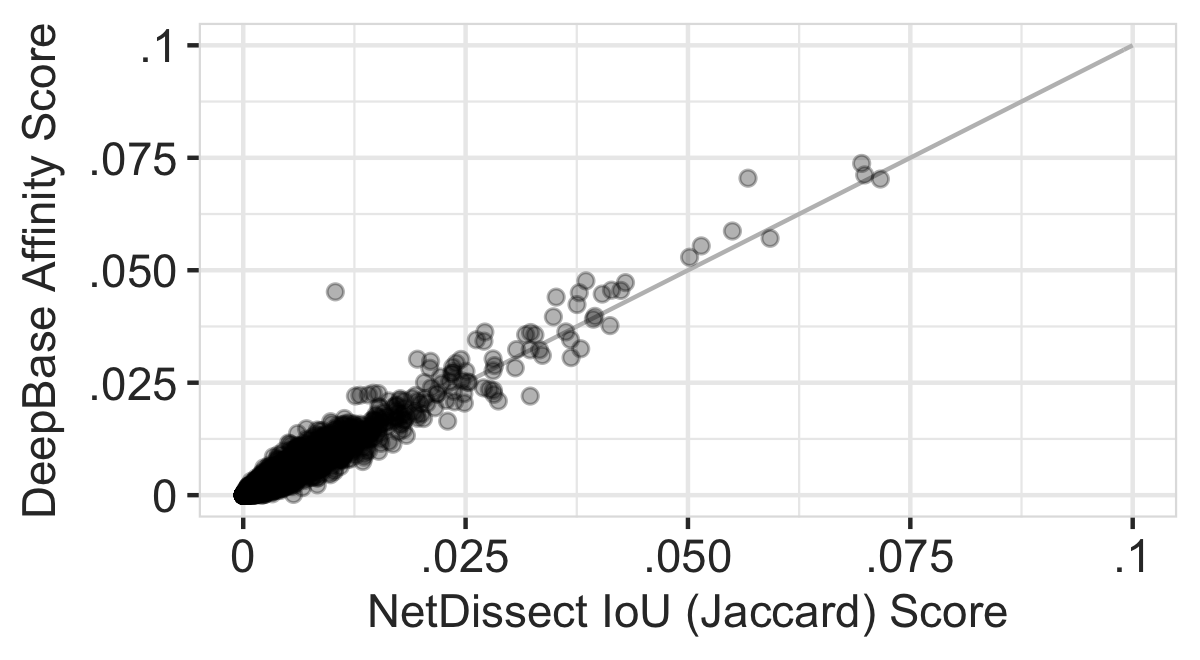}
  \caption{\small Comparison of NetDissect and \sys inspection scores.}
\label{f:exp_netdissect_scores}
\end{figure}
NetDissect is a recent DNI tool developed specifically for CNN models~\cite{bau2017network}. It detects groups of hidden units which act as object and texture detectors (e.g., "the hidden units or layer 2 channel 12 activate specifically for chairs"). To do so, it runs the CNN models on images with pixel-level annotations, and checks which hidden units have activations that correlate with the occurrence of the labels. The affinity measure of the Intersection over Union (i.e., Jaccard similarity), after discretizing the hidden unit activations with quantile-binning.

We replicated this experiment with \sys on over 10K images from a corpus of annotated images provided by NetDissect's authors and designed specifically for this purpose, the Broden dataset\footnote{\url{http://netdissect.csail.mit.edu/data/broden1_227.zip}}. We compare our results with those returned by a public version of NetDissect\footnote{\url{http://netdissect.csail.mit.edu}}. We used a pretrained VGG 16 model, trained on ImageNet data\footnote{\url{http://netdissect.csail.mit.edu/dissect/vgg16_imagenet/}}. Figure~\ref{f:exp_netdissect_scores} presents the result of the analysis. We find that \sys's scores are strongly correlated with NetDissect's, which shows that \sys's declarative interface can express the analysis and the system can produce consistent results. We explain the difference in scores by the fact that several components of the pipeline are non deterministic (among others, the online quantile approximation algorithm and the image up-sampling algorithm used to align the masks with the activations) and environment differences.

\end{document}